\documentclass[review]{elsarticle}

\usepackage{graphicx}
\usepackage{subfigure}
\graphicspath{{./figures/}}
\usepackage{amssymb} 
\usepackage{amsmath} 
\usepackage{physics}
\usepackage{xcolor}
\usepackage{comment}
\usepackage[normalem]{ulem}
\usepackage{delimset}
\usepackage[margin=2.5cm]{geometry}
\usepackage[labelsep=period]{caption}

\usepackage{url}
\usepackage[version=4]{mhchem}
\usepackage[per-mode=symbol]{siunitx}

\usepackage{enumerate}
\usepackage{doi}
\usepackage{hyperref}
\hypersetup{colorlinks=true,linkcolor=blue,citecolor=blue,
			bookmarksnumbered=true,bookmarksopen=true,
			pdfauthor=author,
			pdfpagelayout=SinglePage,pdfstartview=Fit}

\usepackage[capitalise,sort&compress]{cleveref}

\usepackage{booktabs}
\usepackage{multirow}
\usepackage{threeparttable}
\usepackage{setspace}  
\usepackage{array}
\usepackage{makecell}

\usepackage{framed}
\usepackage{multicol}
\usepackage{nomencl}
\makenomenclature
\renewcommand*\nompreamble{\begin{multicols}{2}}
\renewcommand*\nompostamble{\end{multicols}}

\usepackage{microtype}
\usepackage[T1]{fontenc}
\usepackage[utf8]{inputenc}
\usepackage{libertine}
\usepackage[libertine]{newtxmath}
\usepackage{inconsolata}

\biboptions{numbers,sort&compress}
\usepackage{xspace}
\newcommand{\LLItot}{LLI\textsubscript{tot}\xspace}
\newcommand{\LLIcyc}{LLI\textsubscript{cyc}\xspace}
\newcommand{\LAMpe}{LAM\textsubscript{pe}\xspace}

\begin{document}

\begin{frontmatter}

	\title{Degradation model of high-nickel positive electrodes: Effects of loss of active material and cyclable lithium on capacity fade
	}
	\author[ic,faraday]{Mingzhao~Zhuo}\corref{cor}
	\ead{m.zhuo@imperial.ac.uk, mzhuo@connect.ust.hk}
	\author[ic,faraday]{Gregory Offer}
	\ead{gregory.offer@imperial.ac.uk}
	\author[ic,faraday]{Monica Marinescu}
	\ead{monica.marinescu@imperial.ac.uk}
	\address[ic]{Department of Mechanical Engineering, Imperial College London, London, United Kingdom}
	\address[faraday]{The Faraday Institution, United Kingdom}
	\cortext[cor]{Corresponding author}
	\begin{abstract}
		Nickel-rich layered oxides have been widely used as positive electrode materials for high-energy-density lithium-ion batteries, but their degradation has severely affected cell performance, in particular at a high voltage and temperature.
		However, the underlying degradation mechanisms have not been well understood due to the complexity and lack of predictive models.
		Here we present a model at the particle level to describe the structural degradation caused by phase transition in terms of loss of active material (LAM), loss of lithium inventory (LLI), and resistance increase.
		The particle degradation model is then incorporated into a cell-level P2D model to explore the effects of LAM and LLI on capacity fade in cyclic ageing tests.
		It is predicted that the loss of cyclable lithium (trapped in the degraded shell) leads to a shift in the stoichiometry range of the negative electrode but does not directly contribute to the capacity loss, and that the loss of positive electrode active materials dominates the fade of usable cell capacity in discharge.
		The available capacity at a given current rate is further decreased by the additional resistance of the degraded shell layer.
		The change pattern of the state-of-charge curve provides information of more dimensions than the conventional capacity-fade curve, beneficial to the diagnosis of degradation modes.
		The model has been implemented into PyBaMM and made available as open source codes.

	\end{abstract}


\end{frontmatter}


\begin{table*}
	\begin{framed}

		\renewcommand{\nomname}{Acronyms}
		\begin{thenomenclature}

			\nomgroup{A}

			\item [{PE}]\begingroup positive electrode \nomeqref {0}
			\nompageref{1}

			\item [{OCP}]\begingroup electrode open circuit potential \nomeqref {0}
			\nompageref{1}

			\item [{LLI\textsubscript{tot}}]\begingroup loss of total lithium \nomeqref {0}
			\nompageref{1}

			\item [{LAM\textsubscript{pe}}]\begingroup loss of PE active material \nomeqref {0}
			\nompageref{1}

			\item [{DFN}]\begingroup Doyle-Fuller-Newman (or called P2D) \nomeqref {0}
			\nompageref{1}

			\item [{SEI}]\begingroup solid electrolyte interphase \nomeqref {0}
			\nompageref{1}


			\item [{NE}]\begingroup negative electrode \nomeqref {0}
			\nompageref{1}


			\item [{LLI}]\begingroup loss of lithium inventory \nomeqref {0}
			\nompageref{1}

			\item [{LLI\textsubscript{cyc}}]\begingroup loss of cyclable lithium \nomeqref {0}
			\nompageref{1}

			\item [{SoC}]\begingroup state of charge \nomeqref {0}
			\nompageref{1}

			\item [{SPM}]\begingroup single particle model \nomeqref {0}
			\nompageref{1}

			\item [{NMC}]\begingroup nickel manganese cobalt \nomeqref {0}
			\nompageref{1}

		\end{thenomenclature}
		\renewcommand{\nomname}{Nomenclature}
		\begin{thenomenclature}

			\nomgroup{A}

			\item [{$c_\text{p}$}]\begingroup variable of lithium concentration in PE active core \nomeqref {0}
			\nompageref{1}

			\item [{$c_\text{n}$}]\begingroup variable of lithium concentration in NE particle \nomeqref {0}
			\nompageref{1}

			\item [{$c_\text{o}$}]\begingroup variable of lattice-oxygen concentration in PE shell \nomeqref {0}
			\nompageref{1}

			\item [{$s$}]\begingroup variable of core-shell phase boundary location \nomeqref {0}
			\nompageref{1}

			\item [{$c_\text{s}$}]\begingroup fixed concentration of lithium trapped in PE shell \nomeqref {0}
			\nompageref{1}

			\item [{$c_\text{oc}$}]\begingroup fixed concentration of oxygen stored in PE core \nomeqref {0}
			\nompageref{1}

			\item [{$c_\text{e}$}]\begingroup lithium-ion concentration in the electrolyte \nomeqref {0}
			\nompageref{1}

			\item [{$c_\text{thrd}$}]\begingroup threshold value of $c_\text{p}$ for phase transition \nomeqref {0}
			\nompageref{1}

			\item [{$c_\text{p,t}$}]\begingroup top value of $c_\text{p}$ corresponding to 0\% cell SoC \nomeqref {0}
			\nompageref{1}

			\item [{$c_\text{p,b}$}]\begingroup bottom value of $c_\text{p}$ corresponding to 100\% cell SoC \nomeqref {0}
			\nompageref{1}

			\item [{$c_\text{p,max}$}]\begingroup maximum lithium concentration in the PE \nomeqref {0}
			\nompageref{1}


			\item [{$c_{\text{p,surf}}$}]\begingroup surface lithium concentration of PE active core \nomeqref {0}
			\nompageref{1}

			\item [{$c_\text{n,b}$}]\begingroup bottom value of $c_\text{n}$ corresponding to 0\% cell SoC \nomeqref {0}
			\nompageref{1}

			\item [{$c_\text{n,t}$}]\begingroup top value of $c_\text{n}$ corresponding to 100\% cell SoC \nomeqref {0}
			\nompageref{1}

			\item [{$c_\text{n,max}$}]\begingroup maximum lithium concentration in the NE \nomeqref {0}
			\nompageref{1}


			\item [{$c_{\text{n,surf}}$}]\begingroup surface lithium concentration of NE particle \nomeqref {0}
			\nompageref{1}


			\item [{$D_\text{p}$}]\begingroup lithium diffusivity in PE active core \nomeqref {0}
			\nompageref{1}

			\item [{$D_\text{o}$}]\begingroup lattice-oxygen diffusivity in PE shell \nomeqref {0}
			\nompageref{1}

			\item [{$D_\text{n}$}]\begingroup lithium diffusivity in NE particles \nomeqref {0}
			\nompageref{1}

			\item [{$k_{1}$}]\begingroup rate constant of forward phase-transition reaction \nomeqref {0}
			\nompageref{1}

			\item [{$k_{2}$}]\begingroup rate constant of backward phase-transition reaction \nomeqref {0}
			\nompageref{1}

			\item [{$Q$}]\begingroup nominal cell capacity \nomeqref {0}
			\nompageref{1}

			\item [{$I_{\text{app}}$}]\begingroup applied current \nomeqref {0}
			\nompageref{1}

			\item [{$I$}]\begingroup applied current taken by one PE particle on average \nomeqref {0}
			\nompageref{1}

			\item [{$A$}]\begingroup surface area of a single PE particle \nomeqref {0}
			\nompageref{1}

			\item [{$j_\text{p,ave}$}]\begingroup averaged interfacial current density in the PE \nomeqref {0}
			\nompageref{1}

			\item [{$R$}]\begingroup PE particle radius \nomeqref {0}
			\nompageref{1}

			\item [{$a$}]\begingroup active particle surface area per unit electrode volume \nomeqref {0}
			\nompageref{1}

			\item [{$R_\text{shell}$}]\begingroup shell-layer resistance \nomeqref {0}
			\nompageref{1}

			\item [{$\rho$}]\begingroup shell-layer resistivity \nomeqref {0}
			\nompageref{1}

			\item [{$\delta$}]\begingroup shell-layer thickness \nomeqref {0}
			\nompageref{1}

			\item [{$\eta_\text{shell}$}]\begingroup potential drop across the shell layer \nomeqref {0}
			\nompageref{1}

			\item [{$V_\text{a,p}$}]\begingroup active material volume in the PE \nomeqref {0}
			\nompageref{1}

			\item [{$V_\text{a,n}$}]\begingroup active material volume in the NE \nomeqref {0}
			\nompageref{1}

			\item [{$\epsilon_\text{s}$}]\begingroup volume fraction of active materials \nomeqref {0}
			\nompageref{1}

			\item [{$\epsilon_\text{e}$}]\begingroup volume fraction of electrolyte \nomeqref {0}
			\nompageref{1}



			\item [{$M_\text{tot,p}$}]\begingroup total lithium in the PE \nomeqref {0}
			\nompageref{1}

			\item [{$M_\text{tot,n}$}]\begingroup total lithium in the NE \nomeqref {0}
			\nompageref{1}

			\item [{$M_\text{tot}$}]\begingroup total lithium in the PE and NE \nomeqref {0}
			\nompageref{1}

			\item [{$M_\text{tot,0}$}]\begingroup initial total lithium in the PE and NE \nomeqref {0}
			\nompageref{1}

			\item [{$M_\text{cyc,p}$}]\begingroup cyclable lithium in the PE \nomeqref {0}
			\nompageref{1}

			\item [{$M_\text{cyc,n}$}]\begingroup cyclable lithium in the NE \nomeqref {0}
			\nompageref{1}

			\item [{$M_\text{cyc}$}]\begingroup cyclable lithium in the PE and NE \nomeqref {0}
			\nompageref{1}

			\item [{$M_\text{cyc,0}$}]\begingroup initial cyclable lithium in the PE and NE \nomeqref {0}
			\nompageref{1}

			\item [{$j$}]\begingroup interfacial current density \nomeqref {0}
			\nompageref{1}

			\item [{$j_\text{0}$}]\begingroup exchange current density \nomeqref {0}
			\nompageref{1}

			\item [{$k$}]\begingroup rate constant of the charge transfer reaction \nomeqref {0}
			\nompageref{1}


			\item [{$\phi_{\text{s}}$}]\begingroup electric potential of the solid phase \nomeqref {0}
			\nompageref{1}

			\item [{$\phi_{\text{e}}$}]\begingroup electrolyte potential \nomeqref {0}
			\nompageref{1}

			\item [{$\phi_{\text{e,p}}^{\prime}$}]\begingroup electrolyte potential at PE core-shell boundary \nomeqref {0}
			\nompageref{1}

			\item [{$U_{\text{ocp}}$}]\begingroup electrode equilibrium potential \nomeqref {0}
			\nompageref{1}

			\item [{$\sigma$ ($\sigma_\text{eff}$)}]\begingroup bulk (effective) electronic conductivity \nomeqref {0}
			\nompageref{1}

			\item [{$D_\text{e}$ ($D_\text{e,eff}$)}]\begingroup bulk (effective) diffusivity in the electrolyte \nomeqref {0}
			\nompageref{1}

			\item [{$\kappa$ ($\kappa_\text{eff}$)}]\begingroup bulk (effective) ionic conductivity \nomeqref {0}
			\nompageref{1}

			\item [{$\kappa_\text{D,eff}$}]\begingroup effective diffusional ionic conductivity \nomeqref {0}
			\nompageref{1}

			\item [{$f_\text{e}$}]\begingroup mean molar activity coefficient \nomeqref {0}
			\nompageref{1}

			\item [{$t_\text{e}$}]\begingroup transference number \nomeqref {0}
			\nompageref{1}

			\item [{$\alpha$}]\begingroup Bruggeman exponent \nomeqref {0}
			\nompageref{1}

			\item [{$ F $}]\begingroup Faraday constant \nomeqref {0}
			\nompageref{1}

			\item [{$ R $}]\begingroup gas constant \nomeqref {0}
			\nompageref{1}

			\item [{$ T $}]\begingroup absolute temperature \nomeqref {0}
			\nompageref{1}

		\end{thenomenclature}

	\end{framed}
\end{table*}



%
\section{Introduction}

The pursuit of high energy density has driven the widespread application of layered lithium nickel manganese cobalt (NMC) oxides as positive electrode (PE) materials~\cite{Ko2020} of lithium ion batteries, especially those with high nickel ratio such as NMC811.
However, nickel-rich PEs have been shown to suffer from fast capacity decay and low cycling stability due to a multitude of degradation phenomena, among which a major one is the phase transition from the layered structures to disordered spinel and finally to rock-salt structures at low degrees of lithiation~\cite{Edge2021}.
In this work, we develop a degradation model to account for the PE phase transition and embed it into the Doyle-Fuller-Newman (DFN) model within PyBaMM~\cite{Sulzer2021} to quantify the effects of loss of active materials, loss of cyclable lithium, and resistance increase on cell-capacity fade.

Two types of degradation mechanisms have been widely recognized and reported in the literature: structural and chemical decomposition~\cite{Radin2017}.
The first is the structural change through irreversible phase transition of the layered NMC oxides to disordered spinel and rocksalt phases~\cite{Lin2014,Li2021} that do not have the ability to reversibly intercalate lithium.
The phase transition is driven by the fact that the spinel and rocksalt phases are thermodynamically more stable than deintercalated layered oxides~\cite{Radin2017}.
This phase transition is often accompanied with the release of lattice oxygen that eventually leads to formation of oxygen gas (\ce{O2})~\cite{Jung2017,Zhang2020}.
The subsequent reaction of \ce{O2} with the electrolyte results in other gas byproducts such as \ce{CO2} and \ce{CO}~\cite{Jung2017}.
The second mechanism of chemical decomposition entails the chemical and electrochemical reactions at the interface between the NMC active material and electrolyte, which is likely to intertwine with the first mechanism of phase transition.

The surface reactions can result in the dissolution of transition metals (TM), such as Ni, Mn and Co, and the formation of a thin surface layer called pSEI (positive solid electrolyte interphase)~\cite{Edge2021}.
The pSEI layer, consisting of several compounds such as TM fluoride (MF2) and TM carbonates, could impede lithium transport and consume cyclable lithium.
The dissolved transition metal ions were found to migrate through the electrolyte and react on the negative electrode (NE) surface, promoting SEI formation at the NE side, thus consuming lithium ions and raising cell impedance \cite{Gilbert2017,Jung2019,Ruff2021}.
Other degradation mechanisms like mechanical fracturing can further impair the performance by exposing newly-formed PE surface to the electrolyte~\cite{Edge2021} and enabling further structural and chemical degradation.

In spite of the light shed by existing experimental studies, the causes of degradation are still not fully understood~\cite{Radin2017} due to the complex intertwining of various mechanisms.
For example, whether the phase transition or the solid-solution reaction causes the degradation remains an open question~\cite{Lai2020}.
Zhang~\cite{Zhang2020} attributed almost all known problems of NCM811 to the release of lattice oxygen occurring in the irreversible phase transition of layered $\to$ spinel $\to$ rocksalt structures and suggested to focus on suppression of oxygen evolution in order to mitigate degradation.
Ko~et~al.~\cite{Ko2020} proposed that the NMC degradation stems from closely-related chemical and structural changes, which cannot
be interpreted independently.
It becomes apparent that a more thorough understanding of the degradation mechanisms in NMC PEs is urgently needed, including further elucidation from continuing experimental studies.

Meanwhile, modeling studies can help test hypotheses and illuminate the way in which these degradation mechanisms lead to cell capacity fade. Quantitative models can also help in the development of new diagnostic methods, such as those based on degradation mode analysis.
Attempts to develop models of PE degradation especially for NMC materials are rare in the literature.
Ghosh et al. \cite{Ghosh2021} have recently proposed a shrinking-core model to describe the structural phase transition and oxygen release followed by diffusion through the passivation layer.
This model predicts capacity fade during charge by allowing the passivation layer to grow.
Lin~et~al.~\cite{Lin2013} considered manganese dissolution, electrolyte oxidation, and salt decomposition on the PE and SEI growth and manganese deposition on the NE.
Jana~et~al.~\cite{Jana2019} considered electrolyte oxidation and chemo-mechanically induced fracture at the PE side.

This paper aims to model the degradation mechanism of phase transition in the PE and its effect on cyclic capacity fade.
The irreversible phase transition from the layered to rock-salt structures is universally recognized as an inevitable degradation mechanism \cite{Li2019} especially for nickel-rich layered materials, and it seems to be the most important one because of two observations.
Firstly, the electrolyte oxidation and chemical reactions (including pSEI) mainly occur at the beginning of life, much like the SEI formation at the NE side~\cite{Edge2021}, while the phase change and oxygen evolution occur throughout the whole life of the battery~\cite{Zhang2020}.
Secondly, the spinel/rocksalt layer has been reported to have thickness of 15--100 \si{\nano\meter} for a variety of battery chemistries under different usage and storage conditions \cite{Sasaki_2009,Jung2014}, while a surface layer of up to 10 \si{\nano\meter} \cite{Jung2018} was found to build up on the NMC811 PE, which is believed to be the pSEI.

Following the shrinking-core idea to model the phase transition in Ghosh et al.~\cite{Ghosh2021}, we build a particle degradation model (\cref{sec:particle_deg}) by considering the PE degradation as a progressive growth of rocksalt shell from the particle surface to the center (\cref{fig:schematic_deg}) in every single PE particle. 
Compared to Ghosh et al. \cite{Ghosh2021}, we present two improvements. 
Firstly, the particle degradation model is implemented within the DFN model, thus enabling us to account for intermediate/high currents and degradation inhomogeneity across the electrode thickness (\cref{sec:inhomogeneity}). 
Secondly, we explicitly define the primary degradation modes---loss of active material (LAM), loss of lithium inventory (LLI), and resistance increase---and link them to the phase-transition mechanism. 
Specifically, the shell layer formed by the phase transition is assumed to constitute the LAM and trap some cyclable lithium that leads to LLI (\cref{sec:lamlli}); it also hinders lithium-ion transport and thus increases the effective cell resistance (\cref{sec:shellresis}) in a similar way to how the SEI layer is often modeled~\cite{Safari2009}.
The primary degradation modes caused by the phase transition are illustrated in \cref{sec:storage} for the purpose of model verification.
We then demonstrate the model capability of reproducing experimentally-observed phenomena in \cref{sec:cyclicaging}, focusing on the effects of LAM and LLI on cyclic cell-capacity fade.

%

\section{Degradation model of a single particle}
\label{sec:particle_deg}

\begin{figure}
	\centering
	\includegraphics{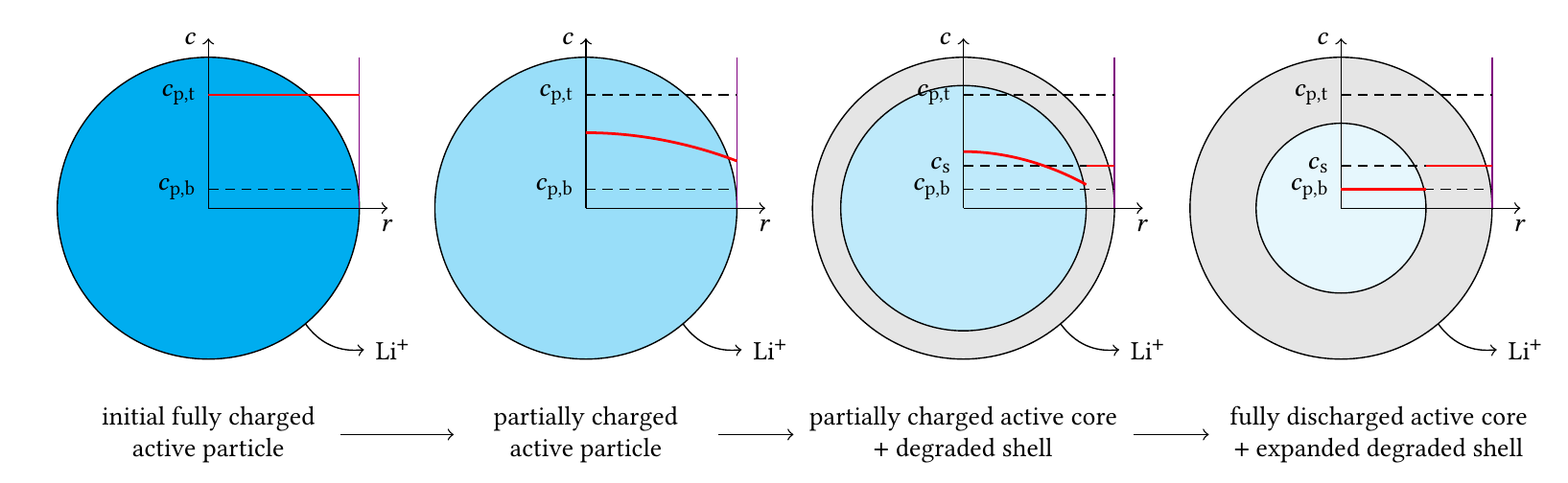}
	\caption{
		Schematic showing degradation of a PE particle under delithiation (i.e., cell charge).
		The degradation advances by the displacement of the phase boundary towards shrinking the core and growing the shell.
		The active core concentration at 0\% cell SoC, the core concentration at 100\% cell SoC, and the fixed concentration of lithium trapped in the degraded shell are denoted by $c_\text{p,t}$, $c_\text{p,b}$, and $c_\text{s}$.
		This schematic is inspired by Ref.~\cite{Zhang2007} in terms of demonstration of the delithiation process.
	}
	\label{fig:schematic_deg}
\end{figure}

This work models the degradation mechanism of phase transition in the PE at extremely low state of lithiation, equivalent to high cell state of charge (SoC).
In this section, we describe a shrinking-core degradation model for a single PE particle; the particle degradation model is then embedded into the DFN model in \cref{sec:dfn} to simulate the cell performance.
The spinel/rocksalt phase is assumed to take the form of a shell and expand towards the center of the particle, via a core-shell moving phase boundary.
The PE delithiation during cell charge is sketched in \cref{fig:schematic_deg} with the phase transition occurring at low stoichiometries, in line with experimental observations~\cite{Bak2014}.
Model \emph{assumptions} are as follows.
\begin{itemize}
	\item The core denotes the active material remaining the properties at the beginning of life and thus allowing lithium to (de)intercalate within the concentration (stoichiometry) range of $c_\text{p,b}$ and $c_\text{p,t}$ in \cref{fig:schematic_deg}.
	\item The degraded shell represents the spinel/rocksalt phase, and a fixed amount of lithium ($c_\text{s}$ in \cref{fig:schematic_deg}) gets trapped in the shell; the shell also serves as a lithium-ion conductor with resistance, leading to an overpotential across the shell.
	\item The (de)intercalation chemical reaction occurs at the core-shell phase boundary with lithium ions supplied through the shell.
	\item The shell-layer growth rate (inwards moving of the phase boundary) depends on the intercalated lithium concentration and on the lattice-oxygen concentration local to the phase boundary, as shown in \cref{eq:pbmoving}.
	\item The degradation is quantified by LAM, LLI, and increase of the shell-layer resistance.
\end{itemize}
Based on these assumptions, the governing equations as well as the boundary conditions are outlined in \cref{sec:governeqs} for a single PE particle; the mass conservation of lithium across the moving phase boundary is used to define the moving-boundary condition.
The LAM and LLI induced by the phase transition are defined and calculated in \cref{sec:lamlli}, and the corresponding shell-layer resistance and overpotential are discussed in \cref{sec:shellresis}.

\subsection{Governing equations and boundary conditions}
\label{sec:governeqs}

In the active core of a PE particle, we consider the diffusion of lithium.
As phase transition proceeds at the core-shell phase boundary, lattice oxygen is released and then diffuses out through the shell.
The governing equations for lithium diffusion in the core and lattice-oxygen diffusion in the shell are
\begin{subequations}
	\begin{align}
		\pdv{c_\text{p}}{t} + \div{\vb{h_{\text{p}}}} & = 0,  \label{eq:corediff}  \\
		\pdv{c_\text{o}}{t} + \div{\vb{h_{\text{o}}}} & = 0,  \label{eq:shelldiff}
	\end{align}
\end{subequations}
where the mass fluxes of lithium and oxygen are expressed as $\vb{h}_{\text{p}} = - D_{\text{p}} \grad{c_\text{p}}$ and $\vb{h}_{\text{o}} = - D_{\text{o}} \grad{c_\text{o}}$, respectively.
At the particle center, the radial symmetry leads to the null-flux boundary condition
\begin{align} \label{eq:center_Neumann_bc}
	\vb{h}_{\text{p}}|_{r=0} = 0
\end{align}
for the lithium diffusion equation.
At the shell outer surface, the oxygen concentration is zero as we assume any oxygen species present there reacts fast:
\begin{align}
	c_\text{o}|_{r=R} = 0.
\end{align}

\begin{figure}
	\centering
	\includegraphics{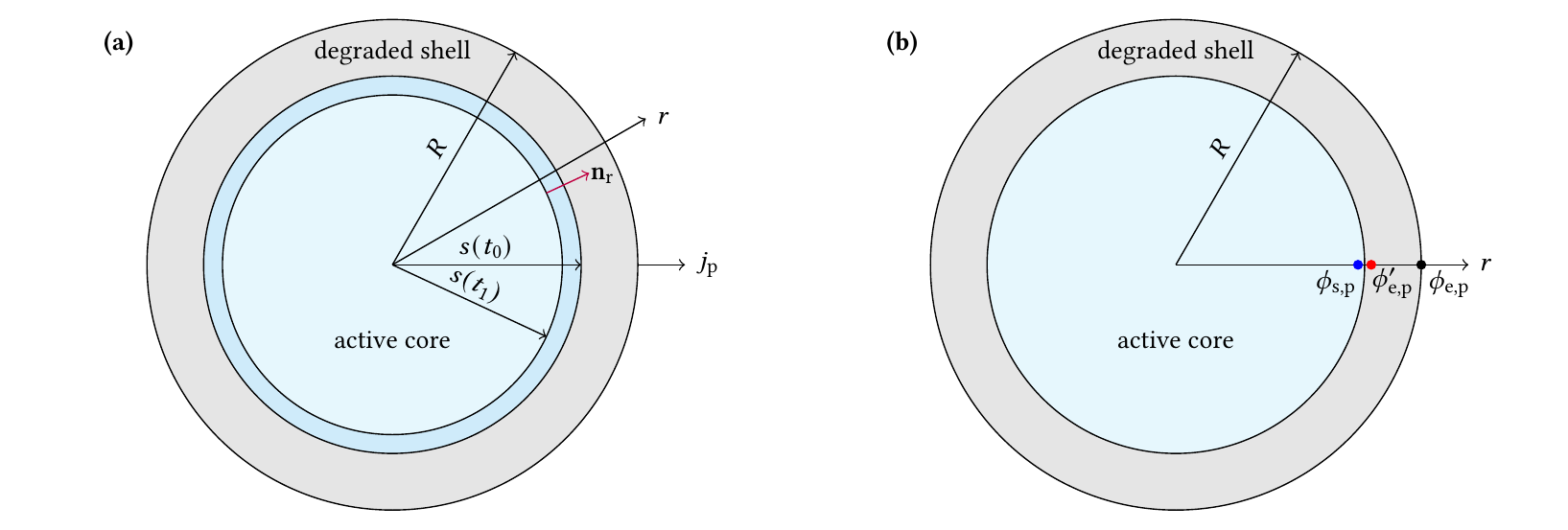}
	\caption{
		Schematic of the shrinking-core degradation model for a PE particle.
		The core represents the active material while the shell denotes the degraded surface layer due to phase transition.
		(a) The displacement of the core-shell phase boundary within a small time interval $\Delta t = t_1 - t_0$ from $r = s \qty(t_0)$ to $r = s \qty(t_1)$, associated with a thin layer of active materials transformed into rocksalt shell.
		$\vb{n}_\text{r}$ denotes the unit vector along the $r$ direction, and $j_\text{p}$ represents the interfacial current density leaving the particle.
		(b) Illustration of the shell-layer overpotential and the potentials used in Butler-Volmer equation: 
		$\phi_\text{s,p}$ (blue point) is the electric potential of the electronically-conductive solid phase, $\phi_{\text{e,p}}^{\prime}$ (red point) is the electrolyte potential at the core-shell phase boundary, and $\phi_\text{e,p}$ (black point) is the electrolyte potential at the shell outer surface.
	}
	\label{fig:core_shell_sche}
\end{figure}

The boundary condition for the lithium diffusion equation~(\ref{eq:corediff}) at the core-shell phase boundary is derived from lithium conservation across the moving phase boundary as follows.
At the core-shell phase boundary, the concentration on the shell side remains constant at $c_{\text{s}}$ as assumed, while the concentration on the core side is the time-dependent variable $c_\text{p}$ that is undetermined.
Consider Fig.~\ref{fig:core_shell_sche}a where a thin layer of fresh phase transition is sketched.
Recall that the core-shell phase boundary moves towards the particle center with time: at time $t = t_0$, the phase boundary is located at $r = s \qty(t_0)$; later at $t = t_1$, it moves to $r = s \qty(t_1)$.
During the small time interval $\Delta t = t_1 - t_0$, a thin layer in the particle, $s \qty(t_1) < r < s \qty(t_0)$, undergoes phase transition from the active core to degraded shell.
Meanwhile, the increase of lithium concentration in this layer is $\Delta c \qty(t) = c_{\text{s}} - c_\text{p} \qty(t)$, and the total lithium increase is
\begin{align} \label{eq:lithium_increase}
	\Delta M = \int_{t_0}^{t_1} 4\pi s^{2}\qty(t) \qty[-\dot{s}\qty(t)] \Delta c \qty(t) \dd{t},
\end{align}
where the time rate of phase-boundary location $\dot{s}\qty(t)$ indicates the speed of the phase transition.
Here we introduce the negative sign in front of $\dot{s}\qty(t)$ to ensure a positive volume of the thin layer.
According to mass conservation, the increase in total lithium increase \cref{eq:lithium_increase} is equal to the amount of lithium that flows into the layer from both the core and the shell:
\begin{align} \label{eq:interface_masscons_1}
	\Delta M = \int_{t_0}^{t_1} 4\pi s^{2}\qty(t) \qty[\vb{h}_{\text{p}} \cdot \vb{n}_\text{r} + \vb{h}_{\text{s}} \cdot (-\vb{n}_\text{r})] \dd{t},
\end{align}
where $\vb{n}_\text{r}$ denotes the unit vector along the $r$ direction at the core-shell phase boundary (\cref{fig:core_shell_sche}a).
Substituting \cref{eq:lithium_increase} into \cref{eq:interface_masscons_1}, we have
\begin{align} \label{eq:interface_masscons_2}
	\int_{t_0}^{t_1} 4\pi s^{2}\qty(t) \qty[
	\dot{s}\qty(t) \Delta c \qty(t) +
	\vb{h}_{\text{p}} \cdot \vb{n}_\text{r} -
	\vb{h}_{\text{s}} \cdot \vb{n}_\text{r}
	] \dd{t} = 0.
\end{align}
Because the above equation holds for all intervals of integration $[t_0, t_1]$, it follows that the integrand must vanish identically:
\begin{align} \label{eq:interface_masscons_3}
	\dot{s} \qty(c_{\text{s}} - c_\text{p}) +
	\vb{h}_{\text{p}} \cdot \vb{n}_\text{r} -
	\vb{h}_{\text{s}} \cdot \vb{n}_\text{r} = 0.
\end{align}

In \cref{eq:interface_masscons_3}, the flux term from the core side can be expressed as
\begin{align} \label{eq:core_heatflux}
	\vb{h}_{\text{p}} \cdot \vb{n}_\text{r} = -D_\text{p} \frac{\partial c_\text{p}}{\partial r} \vb{n}_\text{r} \cdot \vb{n}_\text{r}
	= -D_\text{p} \frac{\partial c_\text{p}}{\partial r},
\end{align}
and the flux term associated with the shell phase is related to the interfacial current density according to mass conservation in the shell:
\begin{align} \label{eq:shell_heatflux}
	\vb{h}_{\text{s}} \cdot \vb{n}_\text{r}   = \qty(R/s)^2 \frac{j_\text{p}}{F},
\end{align}
where the interfacial current density $j_\text{p}$ is positive when lithium leaves the PE particle and $F$ is Faraday constant.
Inserting \cref{eq:core_heatflux,eq:shell_heatflux} into \cref{eq:interface_masscons_3}, we arrive at
\begin{align} \label{eq:interface_masscons_4}
	\dot{s} \qty(c_{\text{s}} - c_\text{p}) -
	D_\text{p} \frac{\partial c_\text{p}}{\partial r} -
	\qty(R/s)^2 \frac{j_\text{p}}{F} = 0.
\end{align}

Following the same procedures, we can obtain the boundary condition for \cref{eq:shelldiff} at the core-shell phase boundary via lattice-oxygen conservation across the moving phase boundary.
Analogous to \cref{eq:interface_masscons_3}, for oxygen conservation we have
\begin{align} \label{eq:interface_oxygen}
	\dot{s} \qty(c_\text{o} - c_{\text{oc}}) -
	\vb{h}_{\text{o}} \cdot \vb{n}_\text{r} = 0,
\end{align}
where $c_{\text{oc}}$ is the fixed concentration of oxygen in the core stored in the form of oxide compounds, and $c_\text{o}$ is the time-dependent variable at the phase boundary on the shell side.
This relation suggests that the phase transition and chemical reactions release the stored oxygen in the core into lattice oxygen that will diffuse out through the shell.
Applying the same operation as in \cref{eq:core_heatflux}, we can reformulate $\vb{h}_{\text{o}} \cdot \vb{n}_\text{r}$ and rewrite \cref{eq:interface_oxygen} as
\begin{align} \label{eq:interface_oxygen_2}
	\dot{s} \qty(c_\text{o} - c_{\text{oc}}) +
	D_\text{o} \frac{\partial c_\text{o}}{\partial r} = 0.
\end{align}
We remark that the time-dependent variables $c_\text{p}$ and $c_\text{o}$ are unknowns at $r=s\qty(t)$, and thus the two boundary conditions~(\ref{eq:interface_masscons_4}) and (\ref{eq:interface_oxygen_2}) are of Robin type, expressing mass conservation of lithium and oxygen across the moving core-shell phase boundary.

To complete \cref{eq:interface_masscons_4,eq:interface_oxygen_2}, the speed of phase transition $\dot{s}$ needs to be specified.
The inwards moving of the phase boundary intrinsically depends on the lithium and oxygen concentrations at the boundary, as well as on the temperature.
The ambient temperature has been observed to accelerate degradation \cite{Jung_2018}.
Due to a distinct lack of experimental data for any validation, we ignore the the temperature effect and model the degradation progress at an accelerated rate.
In particular, the speed of phase transition is taken from Ghosh et al.~\cite{Ghosh2021}:
\begin{align} \label{eq:pbmoving}
	\dot{s} =
	\begin{cases}
		- (k_{1} - k_{2} c_\text{o}), & \text{if} \, \, c_\text{p} < c_{\text{thrd}}; \\
		0,                              & \text{otherwise,}
	\end{cases}
\end{align}
where $c_\text{p}$ and $c_\text{o}$ take values at $r=s\qty(t)$, $k_{1}$ and $k_{2}$ are the rate constants of the forward and backward reactions, and $c_{\text{thrd}}$ is the threshold value below which phase transition occurs.

\subsection{LAM and LLI}
\label{sec:lamlli}

The degradation mechanism of phase transition in a high-nickel PE leads to loss of PE active material (\LAMpe) and loss of lithium inventory (LLI).
In the current work, we assume the shell phase has no ability to store lithium and thus provides no capacity; instead, it remains as a lithium-ion conductor only.
The \LAMpe is thus defined as the volume fraction of the shell phase over the whole particle:
\begin{align} \label{eq:lam}
	\text{LAM}_\text{pe} = 1 - \frac{V_\text{core}}{V_\text{particle}} = 1 - \qty(\frac{s}{R})^3,
\end{align}
where $V_\text{core} = 4\pi s^3/3$ is the volume of the active core and $V_\text{particle} = 4\pi R^3/3$ denotes the particle volume.

The LLI is defined through the difference between the current lithium content and the initial content in both PE and NE.
By convention, the lithium content refers to the total lithium $M_\text{tot} = M_\text{tot,p} + M_\text{tot,n}$ in the electrodes and is calculated as follows:
\begin{subequations} \label{eq:totoal_lithium}
	\begin{align}
		M_\text{tot,p} & = c_\text{p} \times (1-\text{LAM}_{\text{pe}}) \times V_{\text{a,p}}, \label{eq:totoal_lithium_p} \\
		M_\text{tot,n} & = c_\text{n} \times V_{\text{a,n}},
	\end{align}
\end{subequations}
where $c_\text{p}$ and $c_\text{n}$ represent lithium concentrations in the positive core and the negative particle, respectively, and $V_{\text{a}}$ is the active material volume in the PE (``p'') and NE (``n'').
The LLI in terms of total lithium is then defined as
\begin{align} \label{eq:lli_tot}
	\text{LLI}_\text{tot} = 1 - \frac{M_\text{tot}}{M_\text{tot,0}} = 1 - \frac{M_\text{tot,p} + M_\text{tot,n}}{M_\text{tot,p,0} + M_\text{tot,n,0}},
\end{align}
where the subscript ``0'' indicates the initial state.
As the phase transition proceeds, the degraded shell layer grows and traps more lithium that cannot be used in cycling.
Therefore, the total lithium decreases with the phase transition, and thus $\text{LLI}_\text{tot}$ is always nonzero if the phase transition occurs.

\begin{figure}
	\centering
	\includegraphics{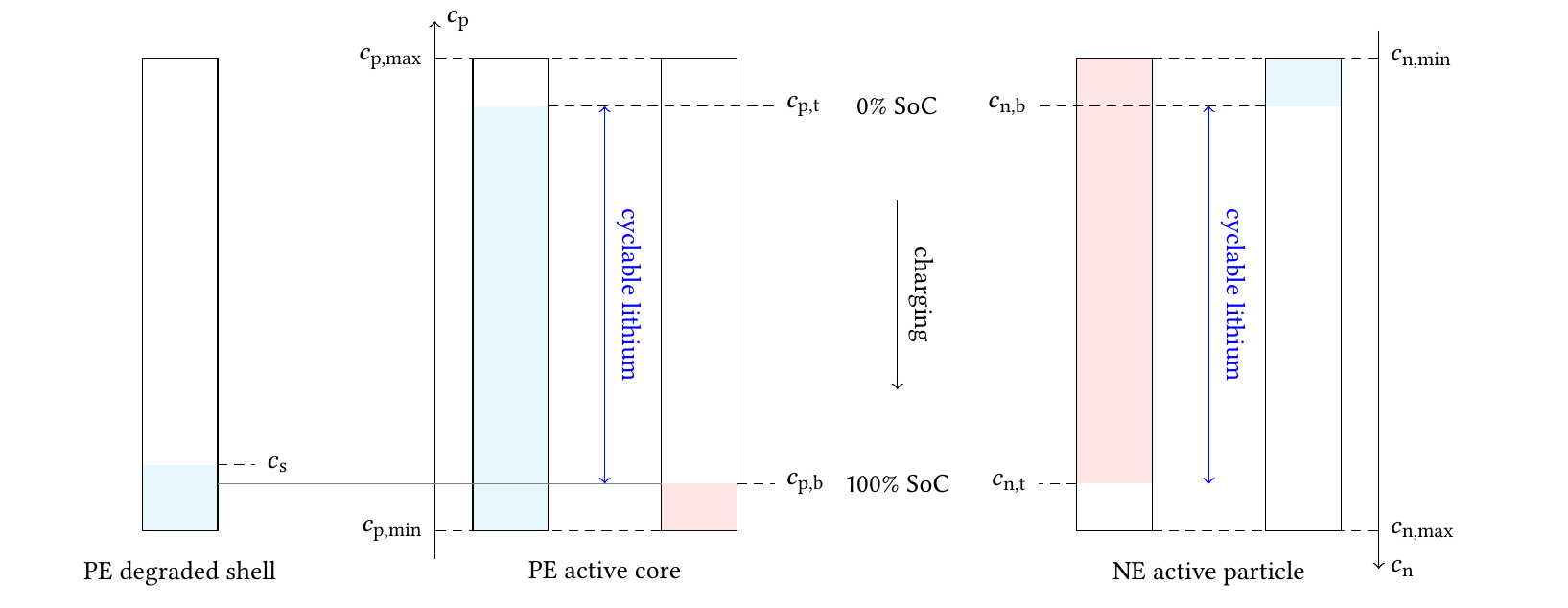}
	\caption{
		Illustration for the cyclable lithium shutting between the PE and NE.
		With the cell SoC varying between 0 and 100\%, the lithium concentration $c_\text{p}$ in the PE particle varies between two limit values [$c_{\text{p,b}}$, $c_{\text{p,t}}$], and the concentration $c_{\text{n}}$ in the NE particle varies within [$c_{\text{n,b}}$, $c_{\text{n,t}}$].
		The 0\% cell SoC is defined by $c_{\text{p,t}}$ in the PE together with $c_{\text{n,b}}$ in the NE, while the 100\% cell SoC is defined by $c_{\text{p,b}}$ in the PE with $c_{\text{n,t}}$ in the NE.
		The concentration of trapped lithium in the degraded shell layer is constant at $c_{\text{s}}$, which is comparable to $c_{\text{p,b}}$.
	}
	\label{fig:lli}
\end{figure}

Note that not all the lithium stored in the active materials shuttle between the PE and NE.
As shown in \cref{fig:lli}, the experimental operating protocol determines the 0\% and 100\% cell SoCs, and the difference between them represents the cyclable lithium.
At the beginning of life, the top value $c_\text{p,t}$ of lithium concentration in the PE paired with the bottom value $c_\text{n,b}$ of concentration in the NE (shaded in light blue) yields the condition for 0\% SoC, while the bottom value $c_\text{p,b}$ in the PE paired with the top value $c_\text{n,t}$ in the NE (light pink) yields the condition for 100\% SoC.
The amount of lithium below $c_\text{p,b}$ (also $c_\text{n,b}$) is not cycled between the PE and NE due to the experimental protocol.
Thus, the loss of this non-cyclable lithium into dead lithium upon shell formation has no impact on the cell performance.
Only the loss of cyclable lithium translates into reduced capacity of the cell.
We further define the total cyclable lithium, based on the 0\% and 100\% SoCs, as
\begin{subequations} \label{eq:totoal_cyclable_lithium}
	\begin{align}
		M_\text{cyc,p} & = (c_\text{p} - c_\text{p,b}) \times (1-\text{LAM}_{\text{pe}}) \times V_{\text{a,p}}, \label{eq:licyc_pe} \\
		M_\text{cyc,n} & = (c_\text{n} - c_\text{n,b}) \times V_{\text{a,n}},
	\end{align}
\end{subequations}
where the bottom limit value $c_\text{p,b}$ ($c_\text{n,b}$) is deducted from the real-time concentration $c_\text{p}$ ($c_\text{n}$), as this amount of lithium is not cycled. 
The corresponding LLI in terms of total cyclable lithium is
\begin{align} \label{eq:lli_cyc}
	\text{LLI}_\text{cyc} = 1 - \frac{M_\text{cyc}}{M_\text{cyc,0}} = 1 - \frac{M_\text{cyc,p} + M_\text{cyc,n}}{M_\text{cyc,p,0} + M_\text{cyc,n,0}}.
\end{align}
Note that the top and bottom values are different from the maximum and minimum concentrations that are determined by the active material itself rather than experimental protocols.

The transition to spinel/rocksalt phase has been observed to occur predominantly when the lithium concentration in the PE is low (see \cref{eq:pbmoving}).
Therefore, the amount of lithium trapped in the shell, denoted as $c_{\text{s}}$, is also expected to be low.
If $c_{\text{s}}$ equals $c_{\text{p,b}}$, we infer that the $\text{LLI}_\text{cyc}$ in \cref{eq:lli_cyc} will be zero, although $\text{LLI}_\text{tot}$ in \cref{eq:lli_tot} is positive.
Because of no loss of cyclable lithium, the cell performance is not affected.
If $c_{\text{s}}$ is not equal to $c_{\text{p,b}}$, the cyclable lithium may increase or decrease, resulting in negative or positive \LLIcyc, respectively.

\subsection{Shell resistance and overpotential}
\label{sec:shellresis}

As the shell is assumed not to store lithium, the intercalation reaction occurs at the core-shell phase boundary.
Meanwhile, the shell layer is assumed to be ion-conductive so that lithium ions can be transferred to the core surface.
This transfer process results in a potential drop across the shell layer, which affects the calculation of overpotential for interfacial current density.
The approach to modeling the shell-layer effects is similar to the conventional treatment of the SEI layer at the NE side~\cite{Safari2009}.

We assume that the resistivity of the shell layer, denoted as $\rho$, is constant and uniform.
Since the shell is generally thin for a working cell, we approximate it as a flat shell with thickness $\delta = R - s$ and surface area $A = 4 \pi R^2$.
The shell layer resistance is formulated as
\begin{align} \label{eq:shellresis}
	R_\text{shell} = \frac{\rho \delta}{A},
\end{align}
and the potential drop across the shell is
\begin{align} \label{eq:shelloverpoten}
	\eta_\text{shell} = I R_\text{shell} = \rho \delta j_\text{p,ave},
\end{align}
where $I$ denotes the averaged current flowing into the electrolyte from a single PE particle and $j_\text{p,ave} = I/A$ is the averaged interfacial current density.

By convention, we use the Butler-Volmer equation to describe the intercalation reaction at the PE active core surface, and the interfacial current density $j_\text{p}$ is expressed as
\begin{align} \label{eq:butlervolmer}
	j_\text{p} =
	j_\text{0,p} \qty[\exp \pqty{\frac{ 0.5 F }{RT} \eta_\text{p} } - \exp \qty(- \frac{ 0.5 F }{RT} \eta_\text{p})],
\end{align}
where~$j_\text{0,p}$ is the exchange current density and~$\eta_\text{p}$ is the reaction overpotential.
The current density~$j_\text{p}$ is defined with a positive value when lithium leaves the PE active material particles.
The exchange current density~$j_0$ is given by
\begin{align}
	j_\text{0,p} = k F c_{\text{p,surf}}^{0.5} \pqty{c_{\text{p,max}} - c_{\text{p,surf}}}^{0.5} c_{\text{e}}^{0.5},
\end{align}
where~$k$~$\qty(\si{\meter^{2.5}\mole^{-0.5}\per\second})$ is the rate constant of the charge transfer reaction, $c_{\text{p,max}}$ is the maximum saturation concentration of intercalated lithium, $c_{\text{p,surf}}$ is the surface lithium concentration of the active core, and $c_{\text{e}}$ is the lithium ion concentration in the electrolyte.

Referring to \cref{fig:core_shell_sche}b, the reaction overpotential~$\eta_\text{p}$ at the PE in \cref{eq:butlervolmer} is expressed, by definition, as
\begin{equation*}
	\eta_\text{p} = \phi_{\text{s,p}} - \phi_{\text{e,p}}^{\prime} - U_{\text{ocp,p}},
\end{equation*}
where the PE equilibrium potential~$U_{\text{ocp,p}}$ is a function of the surface concentration~$c_{\text{p,surf}}$ of the active core, and $\phi_{\text{s,p}}$ is the electric potential of the solid phase in the PE.
Following the approach of Safari et al.~\cite{Safari2009} when addressing the SEI layer overpotential, we relate the electrolyte potential $\phi_{\text{e}}^{\prime}$ at the core-shell phase boundary to the electrolyte potential $\phi_{\text{e}}$ at the shell outer surface via
\begin{align}
	\phi_{\text{e,p}}^{\prime} = \phi_{\text{e,p}} + \eta_\text{shell}.
\end{align}
Thus, the PE reaction overpotential~$\eta_\text{p}$ in \cref{eq:butlervolmer} can now be expressed as
\begin{equation} \label{eq:overpotentail}
	\eta_\text{p} = \phi_{\text{s,p}} - \phi_{\text{e,p}} - \eta_\text{shell} - U_{\text{ocp,p}}.
\end{equation}

\section{Full-cell model}
\label{sec:dfn}

The particle degradation model developed in Section~\ref{sec:particle_deg} describes a single particle in the PE.
In order to estimate the cell degradation, this model is then embedded into a cell-level model such as the DFN model.
The remaining equations of the DFN model are outlined in the following (\cref{sec:NEdiff,sec:cellconseqs}) for completeness.
For more details of the DFN model, readers are referred to the literature~\cite{Cai2011,Marquis2019}.

\subsection{Lithium diffusion in negative particles}
\label{sec:NEdiff}

We consider no degradation in the NE, and thus fickian diffusion is modeled for each NE particle.
The governing equation reads
\begin{align}
	\pdv{c_{\text{n}}}{t} + \div{\qty(- D_{\text{n}} \grad{c_{\text{n}}})} & = 0, \label{eq:negadiff}
\end{align}
where $D_{\text{n}}$ is the diffusivity of intercalated lithium in negative particles.

Radial symmetry dictates that the same Neumann boundary condition as in \cref{eq:center_Neumann_bc} applies to \cref{eq:negadiff}.
At the particle surface, the following Neumann boundary condition is applied
\begin{align} \label{eq:NE_surf_bc}
	- D_\text{n} \frac{\partial c_{\text{n}}}{\partial r} = j_\text{n} / F,
\end{align}
where the interfacial current density is also obtained through Butler-Volmer equation and its expression is readily available in the literature~\cite{Cai2011,Marquis2019}.

\subsection{Conservation equations at cell level}
\label{sec:cellconseqs}

A cell consists of the NE, separator, and the PE.
At the cell level, both the NE and PE consist of a solid phase with electronic conduction~($\phi_{\text{s}}$) and an electrolyte phase with ionic conduction~($\phi_{\text{e}}$) and lithium-ion transport~($c_{\text{e}}$).
For the separator we only need to model the electrolyte phase in a porous medium.
The conservation of electric charge in the solid phase of the two electrodes results in the relation
\begin{align} \label{eq:chargecons_sld}
	\div{ \pqty{- \sigma_{\text{eff}}\, \grad{\phi_{\text{s}}}} } = - aj,
\end{align}
where~$\sigma_{\text{eff}}$ and~$\phi_{\text{s}}$ denote the effective electronic conductivity and electric potential inside the electrode phase, respectively.
Quantity~$j$ is the interfacial current density from the active particle to the electrolyte (charge sink), and $a$ denotes the active particle surface area per unit electrode volume:
\begin{align} \label{eq:a}
	a = \frac{4 \pi R^2}{4/3 \pi R^3} \epsilon_{\text{s}} = \frac{3 \epsilon_{\text{s}}}{R},
\end{align}
where $\epsilon_{\text{s}}$ denotes the active material volume fraction and $R$ the active particle radius.
The conservation of lithium ions in the electrolyte phase of the whole cell~(two electrodes and separator) can be expressed as
\begin{align} \label{eq:masscons_ele}
	\epsilon_{\text{e}} \pdv{ c_{\text{e}} }{t} + \div{ \qty(- D_{\text{e,eff}}) \, \grad{ c_{\text{e}} }} = \qty(1-t_{\text{e}}) a \frac{j}{F},
\end{align}
where~$c_{\text{e}}$ denotes the lithium-ion concentration, $\epsilon_{\text{e}}$ is the volume fraction of electrolyte phase, $D_{\text{e,eff}}$ is the effective diffusivity in the electrolyte, and $t_{\text{e}}$ is the lithium-ion transference number.
The conservation of electric charge in the electrolyte phase of the whole cell is written as
\begin{align} \label{eq:chargecons_ele}
	\div{ \qty(- \kappa_{\text{eff}} \, \grad{ \phi_{\text{e}} } + \kappa_{\text{D,eff}} \, \grad{\ln{c_{\text{e}}}}) } = aj,
\end{align}
where~$\kappa_{\text{eff}}$ is the effective ionic conductivity, $\phi_{\text{e}}$ is the electric potential in the electrolyte, and $\kappa_{\text{D,eff}}$ is the effective diffusional conductivity expressed as
\begin{align} \label{eq:kappaD}
	\kappa_{\text{D,eff}} = \frac{ 2 RT \kappa_{\text{eff}} }{F}
	\qty(1 + \pdv{ \ln{f_{\text{e}}}}{\ln{c_{\text{e}}}}) \qty(1 - t_{\text{e}}),
\end{align}
where~$T$ is the temperature and~$f_{\text{e}}$ is the mean activity coefficient.
The effective electronic conductivity~$\sigma_{\text{eff}}$, effective electrolyte diffusivity $D_{\text{e,eff}}$, and ionic conductivity~$\kappa_{\text{eff}}$ are related to corresponding bulk properties through the Bruggeman correlation~\cite{Tjaden2016}:
\begin{align*}
	\sigma_{\text{eff}}       = \sigma \epsilon^{\alpha}_{\text{s}},  \quad
	D_{\text{e,eff}} = D_{\text{e}} \epsilon^{\alpha}_{\text{e}},  \quad
	\kappa_{\text{eff}}       = \kappa \epsilon^{\alpha}_{\text{e}},
\end{align*}
where $\alpha$ is the Bruggeman exponent, and $\sigma$, $D_{\text{e}}$, and~$\kappa$ are bulk material properties that can be concentration dependent.

\subsection{Cell voltage}

The DFN model resolves the variation of field variables in the through-cell direction and thus is especially needed for (dis)charges at a high-current rate where the lithium-ion concentration gradient in the electrolyte cannot be ignored.
However, for the cases where averaged quantities over the electrode thickness are sufficient to describe the cell performance, a suitable choice is the SPM.
The same PE phase-transition model as described in the previous section can be embedded into any given SPM, should the latter be preferred.
The fickian diffusion model for the PE particle in the SPM would be replaced by the developed particle degradation model.
The simplification from the DFN model to SPM can be found for example in Marquis et al.~\cite{Marquis2019}, and here we just detail the cell terminal voltage calculation in the SPM for better interpretation of the following simulation results.

Similar to the PE reaction overpotential in \cref{eq:overpotentail}, the NE reaction overpotential is expressed as
\begin{equation} \label{eq:NEoverpotentail}
	\eta_\text{n} = \phi_{\text{s,n}} - \phi_{\text{e,n}} - U_{\text{ocp,n}},
\end{equation}
except for the shell overpotential term $\eta_\text{shell}$.
The difference between the solid-phase potential of the PE and that of the NE is defined as the cell terminal voltage and can be obtained by subtracting the two equations (\ref{eq:NEoverpotentail}) and (\ref{eq:overpotentail}):
\begin{align} \label{eq:terminalvolt}
	V_{\text{t}} = \phi_{\text{s,p}}-\phi_{\text{s,n}} = U_{\text{ocp,p}} \qty(c_{\text{p,surf}}) - U_{\text{ocp,n}} \qty(c_{\text{n,surf}}) + \eta_{\text{p}} - \eta_{\text{n}} + \eta_{\text{shell}}.
\end{align}
Note that in the SPM, we ignore the electrolyte effect and consider equal electrolyte potentials $\phi_{\text{e,p}}=\phi_{\text{e,n}}$ in the PE and NE.

\section{Results and discussion}

This section demonstrates the capability of the proposed model to describe the \LAMpe, \LLIcyc, and shell layer resistance $R_\text{shell}$, as well as their contribution to the cell capacity fade.
We first demonstrate in \cref{sec:storage} that the model performs as expected in the case of calendar ageing.
Then, we run cyclic ageing tests to demonstrate effects of the \LAMpe, \LLIcyc, and $R_\text{shell}$ on the capacity fade in \cref{sec:cyclicaging}.
In the first two sections, we choose the Single Particle Model (SPM in PyBaMM) to accommodate the particle degradation model because the focus
is to analyze the overall response of the electrode at low current rates.
We then plug the particle degradation model into the DFN model (DFN in PyBaMM) to demonstrate degradation inhomogeneity in the electrode thickness direction at medium and high current rates (\cref{sec:inhomogeneity}).

The parameter values for the numerical simulation are taken from Sturm et al.~\cite{Sturm2019} and Ghosh et al.~\cite{Ghosh2021} for a commercial lithium-ion cell (INR18650-MJ1, LG Chem) with NMC811 cathode and SiC anode.
These parameter values can be found in \cref{tab:params}.
\begin{table}
	\centering
	\footnotesize
	\caption{Parameter values for simulation of a commercial lithium-ion cell with NMC811 positive electrode (PE) and SiC negative electrode (NE).}
	\label{tab:params}
	\renewcommand{\arraystretch}{1.3}
	\begin{threeparttable}
		\begin{tabular*}{\textwidth}{@{}l@{\extracolsep{\fill}}lcccl@{}} \toprule
			Parameter    & symbol       & unit       & \multicolumn{2}{c}{value} & note/ref.\\
			\midrule
			&        &        &  PE ($i=\text{p}$) & NE ($i=\text{n}$) &  \\
			\cmidrule (ll){4-5}
			particle radius   & $R_{i}$ & \si{\micro\meter} & 3.8 & 6.1 & \cite{Sturm2019} \\
			active material volume fraction & $\epsilon_{\text{s}}$ & - & 0.745 & 0.694 & \cite{Sturm2019} \\
			electrolyte volume fraction (porosity) & $\epsilon_{\text{e}}$ & - & 0.171 & 0.216 & \cite{Sturm2019} \\
			maximum lithium concentration & $c_{i,\text{max}}$ & \si{\mole\per\cubic\meter} & 49340 & 34257 & \cite{Ghosh2021} \\
			stoichiometry at 0\%\,SoC\tnote{a} & $c_{i}/c_{i,\text{max}}$ & - & 0.942 & 0.002 & \cite{Ghosh2021} \\
			stoichiometry at 100\%\,SoC\tnote{a} & $c_{i}/c_{i,\text{max}}$ & - & 0.222 & 0.852 & \cite{Ghosh2021} \\
			diffusivity & $D_i$ & \si{\square\meter\per\second} & \num{1.0e-14} & \num{1.0e-14} & \cite{Ghosh2021} \\
			charge-transfer reaction rate &  $k$ & \si{\meter^{2.5}\mole^{-0.5}\per\second} &  \num{3.2e-11} & \num{1.0e-11} & \cite{Ghosh2021} \\
			electronic conductivity & $\sigma$ & \si{\square\meter\per\second} & \num{0.17} & \num{100} & \cite{Sturm2019} \\
			electrode thickness & $L_\text{ed}$ & \si{\micro\meter} & $86.7$ & \num{66.2} & \cite{Sturm2019} \\
			open circuit potential & $U_\text{ocp}$ & \si{\volt} & data\tnote{b} & data\tnote{b} & \cite{Sturm2019} \\
			initial concentration & $c_{i,0}$ & \si{\mole\per\cubic\meter} & \num{46478.28} & \num{68.514} & 0\%\,SoC \\
			\cmidrule (ll){4-5}
			initial oxygen concentration in PE shell & $c_\text{o,0}$   &  \si{\mole\per\cubic\meter} & \multicolumn{2}{c}{0} & \\
			initial phase boundary location & $s_{0}$ & \si{\micro\meter}  & \multicolumn{2}{c}{\num{3.75}} &  \\
			threshold value for phase transition & $c_{\text{thrd}}$ & \si{\mole\per\cubic\meter} & \multicolumn{2}{c}{\num{14802}} & 0.3 $c_{\text{p,max}}$  \\
			fixed oxygen concentration in the core & $c_\text{oc}$ & \si{\mole\per\cubic\meter} & \multicolumn{2}{c}{\num{152193.21}} & \cite{Ghosh2021} \\
			oxygen diffusivity in the shell & $D_\text{o}$ & \si{\square\meter\per\second} & \multicolumn{2}{c}{\num{1E-17}} & \cite{Ghosh2021} \\
			forward reaction rate constant & $k_{1}$ & \si{\meter\per\second} & \multicolumn{2}{c}{\num{0.8544E-11}} & \cite{Ghosh2021} \\
			reverse reaction rate constant & $k_{2}$ & \si{\meter^{4}\per\mole\per\second} & \multicolumn{2}{c}{\num{1.732E-16}} & \cite{Ghosh2021} \\
			PE shell resistivity & $\rho$ &  \si{\ohm\meter} & \multicolumn{2}{c}{\num{1e6}} & \cite{Safari2009}\tnote{c} \\
			%
			nominal cell capacity & $Q$ & \si{\ampere\hour}  & \multicolumn{2}{c}{\num{3.35}} & \cite{Sturm2019,Ghosh2021} \\
			lower cut-off voltage & $V_\text{lower}$ & \si{\volt} & \multicolumn{2}{c}{\num{2.8}} & \\
			upper cut-off voltage & $V_\text{upper}$ & \si{\volt} & \multicolumn{2}{c}{\num{4.2}} & \\
			current collector/electrode interface area & $A_\text{cc}$ & \si{\square\meter} & \multicolumn{2}{c}{\num{7.134e-2}} & \cite{Ghosh2021} \\
			lithium-ion concentration in electrolyte & $c_\text{e}$ & \si{\mole\per\cubic\meter} & \multicolumn{2}{c}{\num{1000}} & PyBaMM\tnote{d}   \\
			electrolyte diffusivity          & $D_\text{e}$  & \si{\square\metre\per\second}   & \multicolumn{2}{c}{$\num{5.34e-10}\,e^{-0.65 c_\text{e}}$}                                          & \cite{Capiglia1999} \\
			ionic conductivity   & $\kappa$   & \si{\siemens\per\meter}     & \multicolumn{2}{c}{$\num{0.0911} + 1.9101 c_\text{e} - 1.052 c_\text{e}^2 + 0.1554 c_\text{e} ^3$}        & \cite{Capiglia1999} \\
			transference number  & $t_\text{e}$ & $-$           & \multicolumn{2}{c}{\num{0.4}}       & PyBaMM\tnote{d} \\
			thermodynamic factor & $1 + \partial \ln{f_{\text{e}}} / \partial \ln{c_{\text{e}}}$ & $-$ & \multicolumn{2}{c}{\num{1}} & PyBaMM\tnote{d}            \\
			separator porosity & $\epsilon_\text{e}$ & $-$ & \multicolumn{2}{c}{\num{0.45}}  & PyBaMM\tnote{d} \\
			separator thickness & $L_\text{sep}$ &   \si{\micro\meter} & \multicolumn{2}{c}{\num{12}} & \cite{Sturm2019} \\
			Faraday constant  & $F$ & \si{\coulomb\per\mole}   & \multicolumn{2}{c}{\num{96485}} &  \\
			gas constant    & $R$   & \si{\joule\per\kelvin\per\mole}          & \multicolumn{2}{c}{\num{8.31}} &  \\
			absolute temperature & $T$ & \si{\kelvin}     & \multicolumn{2}{c}{\num{298.15}} &  \\
			Bruggeman exponent & $\alpha$ & $-$     & \multicolumn{2}{c}{\num{1.5}} & \cite{Sturm2019} \\
			\bottomrule
		\end{tabular*}
		\begin{tablenotes}
			\item[a] See \cref{fig:lli}.
			\item[b] Refer to Fig. 2a and b in Sturm et al.~\cite{Sturm2019} for electrode OCP curves and the link therein to download the data.
			\item[c] Estimated based on a SEI layer resistivity.
			\item[d] Open-source repository: \url{https://github.com/pybamm-team/PyBaMM}.
		\end{tablenotes}
	\end{threeparttable}
\end{table}

\subsection{LAM and LLI in calendar ageing}
\label{sec:storage}

In this section, we quantify three degradation modes---the \LAMpe, \LLIcyc, and shell resistance---in the scenario of calendar ageing at a high cell voltage (low level of PE lithiation).
The \LAMpe and shell resistance are directly calculated from \cref{eq:lam,eq:shellresis}, respectively, depending on the shell-layer thickness.
However, the \LLIcyc defined in \cref{eq:lli_cyc} does not show a straightforward dependence on the extent of degradation and thus needs further numerical studies.
We assumed that the lithium present in the degraded shell is fixed at a constant concentration level $c_\text{s}$ and is indefinitely trapped such that it no longer contributes to the cell capacity.
The parameter $c_\text{s}$ is a key factor impacting the calculation of \LLIcyc via the mass conservation across the core-shell phase boundary as expressed in \cref{eq:interface_masscons_4}.
To understand how cyclable lithium is lost, we investigate the effect of the parameter $c_\text{s}$.

The model is run for the following scenario: from fully discharged, the cell is charged at a low constant-current rate (0.5\,C) until the upper cut-off voltage of \SI{4.2}{\volt}; the cell is then stored for 6 hours, enabling degradation to occur.
Three cases regarding the value of $c_\text{s}$ are considered: (I) $c_\text{s} = 0.182\,c_\text{p,max} < c_\text{p,b}$, (II) $c_\text{s} = 0.222\,c_\text{p,max} = c_\text{p,b}$, and (III) $c_\text{s} = 0.324\,c_\text{p,max} > c_\text{p,b}$.

\begin{figure}
	\centering
	\includegraphics{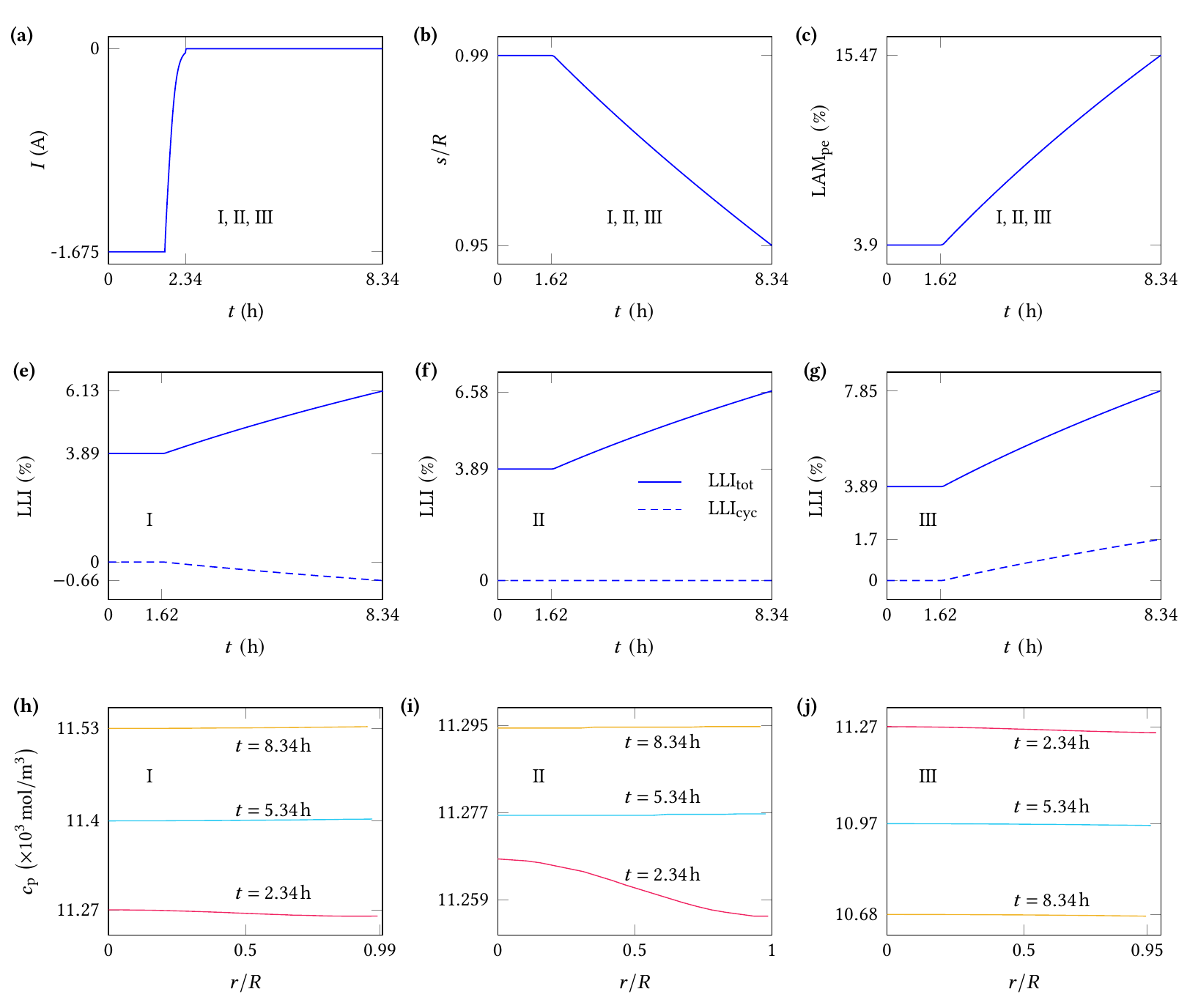}
	\caption{
		Effects of the trapped lithium content $c_\text{s}$ on the loss of total lithium \LLItot and loss of total cyclable lithium \LLIcyc.
		Three cases are considered: (I) $c_\text{s} = 0.182\,c_\text{p,max} < c_\text{p,b}$, (II) $c_\text{s} = 0.222\,c_\text{p,max} = c_\text{p,b}$, and (III) $c_\text{s} = 0.324\,c_\text{p,max} > c_\text{p,b}$.
		The experimental protocol is the same for all three cases, and so is the applied current $I$ (a).
		The phase boundary location $s$ (b) and \LAMpe (c) are the same for all three cases, as they do not depend on $c_\text{s}$.
		The LLI evolution and lithium concentration $c_\text{p}$ at three time instants during the 6-hour rest are plotted for case (I) in subplots (e) and (h), for case (II) in subplots (f) and (i), and for case (III) in subplots (g) and (j).
	}
	\label{fig:res_LLI}
\end{figure}

As shown in \cref{fig:res_LLI}, the storage starts at $t = \SI{2.34}{\hour}$ and ends at $t = \SI{8.34}{\hour}$.
The degradation is triggered at the end of the charge ($t = \SI{1.62}{\hour}$) when the lithium concentration goes below the threshold value according to \cref{eq:pbmoving}.
\cref{fig:res_LLI}a shows the current profile, confirming the protocol specified above.
Since the parameter $c_\text{s}$ does not affect the progress of phase transition, the core-shell phase boundary $s/R$ and the \LAMpe evolve in identical ways for all three cases, as shown in \cref{fig:res_LLI}b--c: they remain at the values of the initial conditions until the PE degradation starts at $t = \SI{1.62}{\hour}$, after which $s/R$ decreases as the shell thickens and the \LAMpe increases accordingly.

The LLI including \LLItot and \LLIcyc are shown in \cref{fig:res_LLI}e--g for all three cases.
Since the lithium trapped in the shell is considered as a pure loss, it is excluded in the total lithium calculation in \cref{eq:totoal_lithium_p} via the deduction of \LAMpe.
Hence, the \LLItot is always positive and increases with degradation.
From case (I) to (III), we can see that the higher the $c_\text{s}$ value, the greater the \LLItot, in spite of the same degradation and shell volume growth for all three cases.
Note that the initial value is the same at 3.89\% because the same amount of lithium loss is considered for the initial shell of nonzero thickness.

In contrast, the \LLIcyc varies depending on the value of $c_\text{s}$ relative to that of $c_\text{p,b}$: it decreases with progressive phase transition when $c_\text{s} < c_\text{p,b}$ (case I), suggesting that extra cyclable lithium is harvested due to the phase transition; when $c_\text{s} > c_\text{p,b}$ (case III), the \LLIcyc is positive and increases with the degradation; if $c_\text{s} = c_\text{p,b}$ (case II), the \LLIcyc remains null, regardless of the phase transition occurring.
The explanation is as follows.
In the core, the lithium below the level $c_\text{p,b}$ (corresponding to 100\% SOC in \cref{fig:lli}) is not cycled and temporarily gets ``trapped'' in the core; only the rest lithium above the concentration $c_\text{p,b}$ is taken into account in the calculation of total cyclable lithium (\cref{eq:licyc_pe}).
Consider the transition of a thin layer of active core to the shell phase (see \cref{fig:core_shell_sche}a).
When $c_\text{s} = c_\text{p,b}$, the ``trapped'' lithium (up to $c_\text{p,b}$) in the thin layer of active core just continues to be trapped in the
shell after the transition, while all the remaining cyclable lithium is pushed to the untransformed active core due to the mass conservation across the phase boundary.
In this process, there is no loss and gain of cyclable lithium.
If $c_\text{s} > c_\text{p,b}$, some cyclable lithium from the active core, including the thin layer, will be consumed to fill the gap between $c_\text{s}$ and $c_\text{p,b}$, leading to positive \LLIcyc.
If $c_\text{s} < c_\text{p,b}$, some part of the ``trapped'' lithium in the thin layer of active core, together with the rest cyclable lithium, is transferred to the remaining active core and thus remains cyclable.
We remark that the harvest of cyclable lithium in case I seems counter-intuitive but does not violate physics.

The variation of lithium concentration $c_\text{p}$ in PE active core is shown in \cref{fig:res_LLI}h--j as the phase transition proceeds.
After the voltage control ($t = \SI{2.34}{\hour}$), the concentration $c_\text{p}$ in the active core is approximately equal to $\SI{11.27}{\kilo\mole\per\cubic\meter}$, which is higher than $c_\text{p,b} = \SI{10.95}{\kilo\mole\per\cubic\meter}$ that corresponds to the 100\% cell SoC.
In cases I and II, the values of $c_\text{s}$ are $\SI{9}{\kilo\mole\per\cubic\meter}$ and $\SI{10.95}{\kilo\mole\per\cubic\meter}$, respectively, both lower than $c_\text{p} = \SI{11.27}{\kilo\mole\per\cubic\meter}$ in the core.
As the phase transition occurs, the difference between $c_\text{p}$ and $c_\text{s}$ is transferred to the untransformed core due to the mass conservation law.
As a result, the lithium concentration $c_\text{p}$ in the untransformed core increases slightly during the phase transition---see the concentration profiles at three time instants in \cref{fig:res_LLI}h and i.
However, in case (III), $c_\text{s} = \SI{16}{\kilo\mole\per\cubic\meter}$ is higher than $c_\text{p} = \SI{11.27}{\kilo\mole\per\cubic\meter}$ in the core.
The lithium is thus transferred from the untransformed core to the newly-formed shell to fill the concentration gap, and the lithium concentration $c_\text{p}$ in the untransformed core decreases accordingly with time as shown in \cref{fig:res_LLI}j.
In conclusion, the lithium concentration in the active core evolves depending on its value relative to the value of $c_\text{s}$, unlike the \LLIcyc.

\subsection{Capacity fade in cyclic ageing}
\label{sec:cyclicaging}


In the previous section, we demonstrated that the phase transition causes the \LAMpe, \LLIcyc, and shell resistance, and in this section we run cyclic tests to explore their effects on cell capacity fade.
The cyclic ageing tests in our simulations are defined to closely resemble experimental ones, and a typical cycle is specified as
\begin{center}
	\begin{minipage}{0.28\linewidth}
		\emph{Charge at 0.5 C until 4.2 V, \\
			Hold at 4.2 V until C/50, \\
			Rest for 60 minutes, \\
			Discharge at 0.5 C until 2.8 V, \\
			Hold at 2.8 V until C/50, \\
			Rest for 60 minutes.}
	\end{minipage}
\end{center}
According to \cref{eq:pbmoving}, the degradation mainly occurs at low lithiation levels in the PE.
The voltage control and rest are thus added after the constant-current charge to allocate more time for degradation.
Also, an accelerated degradation rate, rather than a real-world one, is used to reduce the simulation cost.
A total of 20 consecutive cycles are repeated for three simulated scenarios: (I) only \LAMpe, (II) \LAMpe and \LLIcyc, and (III) \LAMpe, \LLIcyc, and shell resistance.

\begin{figure}
	\centering
	\includegraphics{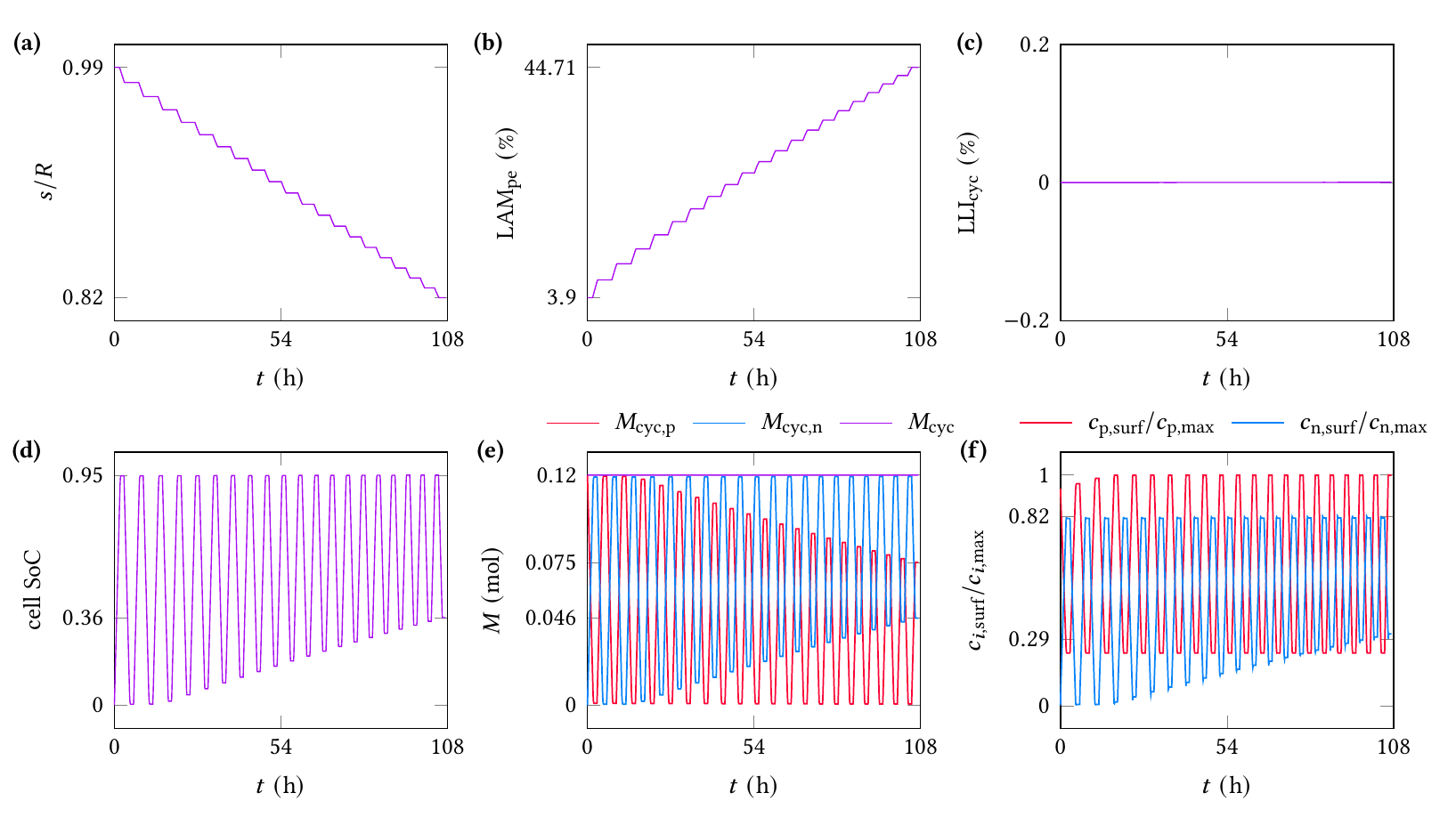}
	\caption{
	Scenario I: effects of loss of PE active material (\LAMpe) on cell performance in cyclic ageing tests when the loss of cyclable lithium (\LLIcyc) is disabled.
	(a) core-shell phase boundary ($s/R$), (e) total cyclable lithium ($M_\text{cyc}$) in both electrodes, cyclable lithium in the PE ($M_\text{cyc,p}$) and NE ($M_\text{cyc,n}$), and (f) normalized particle surface concentration/stoichiometry ($c_{i,\text{surf}}/c_{i,\text{max}}$).
	}
	\label{fig:sce1}
\end{figure}

The results of scenario I are shown in \cref{fig:sce1} to explore the effects of \LAMpe exclusively.
In this case, we set the parameter $c_{\text{s}} = c_{\text{p,b}}$ so that there is no loss of cyclable lithium and set $\rho = 0$ to remove the shell resistance and overpotential across the shell layer.
The calculated \LLIcyc is zero, as can be directly seen in \cref{fig:sce1}c or be cross-checked by the constant total cyclable lithium $M_\text{cyc} = M_\text{cyc,p} + M_\text{cyc,n}$ in both the PE and NE as shown in \cref{fig:sce1}e.

In \cref{fig:sce1}a, the normalized phase boundary location $s/R$ decreases with time, indicating the phase transition occurs towards the particle center as designed.
However, it does not decreases at all times, and this is because of the condition specified in \cref{eq:pbmoving} that the phase transition only occurs when the core surface concentration is lower than the threshold value.
Hence, the phase transition occurs and the phase boundary location decreases mainly at the end of cell charge and in the following voltage hold and rest.
Correspondingly, the \LAMpe increases with time (\cref{fig:sce1}b) in a similar fashion to the phase boundary location, as governed by \cref{eq:lam}.
Note that we specify a nonzero initial shell thickness to avoid numerical issues arising in the transformation of computational domains, e.g., a zero denominator in \cref{eq:varichange}.

\cref{fig:sce1}d shows the evolution of the cell SoC, which is calculated through coulomb counting:
\begin{align} \label{eq:soc}
	\text{SoC} = - \frac{1}{Q} \int_{0}^{t} I_\text{app} \qty(t) \dd{t},
\end{align}
where $Q$ is the nominal capacity and $I_\text{app}$ is the applied current with positive sign on discharge.
As a cycle begins, the SoC first increases during the charge and the following voltage control, and then it stays constant in the rest;
in the latter half of a cycle, the SoC decreases during the discharge and voltage control and then stays unchanged during the rest.
This variation pattern is repeated from cycle to cycle.
However, the SoC upper limit stays unchanged, while the lower limit increases with the cycle number.
The SoC range is thus shrinking, leading to cell capacity fade.

The shrinkage of the SoC range is due to the phase-transition-induced degradation, and the specific change pattern is caused by the increasing \LAMpe and zero \LLIcyc, as explained below.
At the end of charge, all the cyclable lithium is supposed to shuttle into the NE.
In our model, the NE does not degrade and thus can accommodate all the cyclable lithium, regardless of the cycle number.
In this scenario, there is no loss of cyclable lithium; therefore, the SoC upper limit at the end of charge does not vary with cycle number.
This is also confirmed by the unvarying upper limit of cyclable lithium $M_\text{cyc,n}$ in the NE in \cref{fig:sce1}e that equals the total cyclable lithium $M_\text{cyc}$.
Now consider the other side.
As a discharge starts, the cyclable lithium is transferred back to the PE.
However, due to the \LAMpe, the PE is not able to accommodate all the cyclable lithium.
The more severe degradation, the greater \LAMpe, the less lithium the PE can accommodate, and the more lithium left in the NE at the end of discharge.
This is verified by the observation in \cref{fig:sce1}e that the upper limit of cyclable lithium $M_\text{cyc,p}$ in the PE decreases, while the lower limit of cyclable lithium $M_\text{cyc,n}$ in the NE increases.
Therefore, with increasing cycle number and \LAMpe, the lower limit of the SoC is pushed higher.

\begin{figure}
	\centering
	\includegraphics{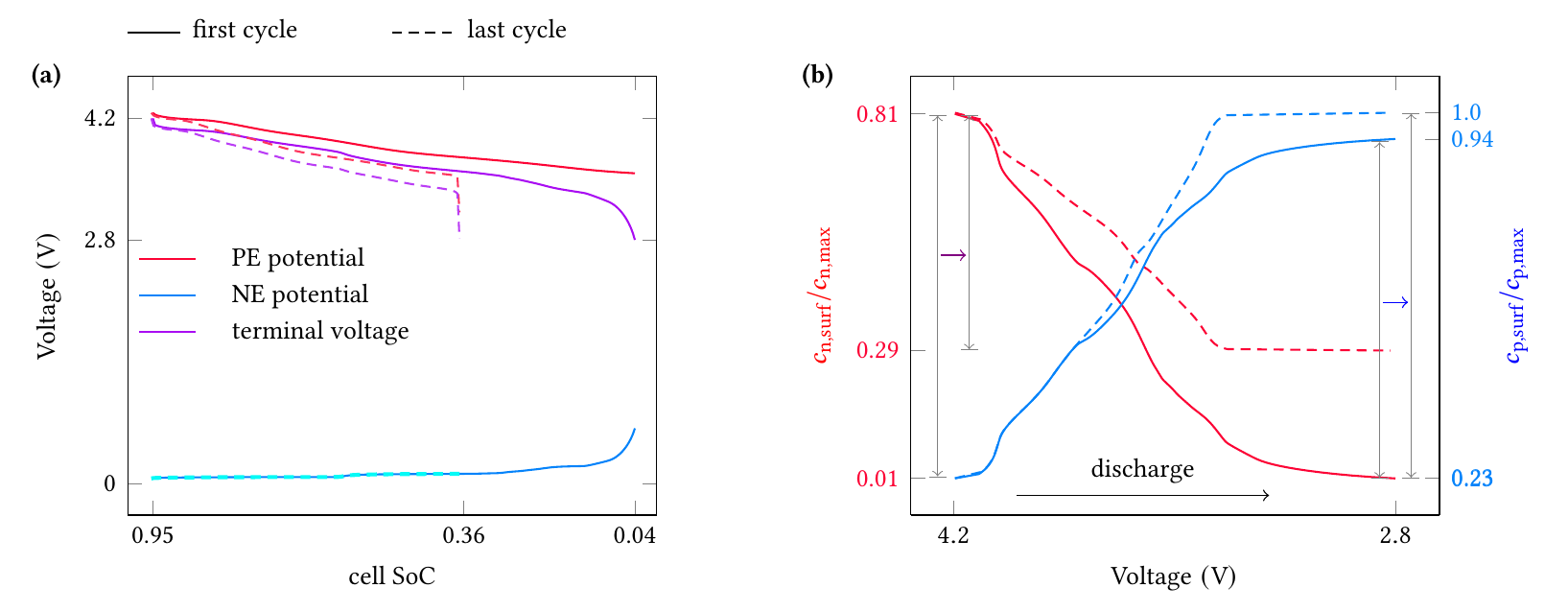}
	\caption{
		Scenario I: (a) terminal voltage, PE potential, and NE potential versus the cell SoC in the discharge step of the first and last cycles; (b) shrinkage of the NE stochiometry range and expansion of the PE stochiometry range during discharge.
	}
	\label{fig:sce1_sto}
\end{figure}

\cref{fig:sce1}f shows the particle surface concentrations ($c_{i,\text{surf}}/c_{i,\text{max}}$) in the PE ($i=\text{p}$) and NE ($i=\text{n}$) normalized by respective maximum concentrations.
Two observations follow.
First, the upper limit of $c_\text{n,surf}/c_\text{n,max}$ basically does not change, and so does the lower limit of $c_\text{p,surf}/c_\text{p,max}$.
Second, in accordance with the increasing lower limit of $M_\text{cyc,n}$ in \cref{fig:sce1}e, the lower limit of $c_\text{n,surf}/c_\text{n,max}$ keeps increasing with cycle number, and this increase drives its counterpart---the upper limit of $c_\text{p,surf}/c_\text{p,max}$ to increase slightly to meet the fixed lower cut-off voltage of \SI{2.8}{\volt} (see the following discussion of \cref{fig:sce1_sto} for more details).
Note that this resultant slight increase in the PE particle does not conflict with the decreasing upper limit of total cyclable lithium $M_\text{cyc,p}$ in \cref{fig:sce1}e because of the \LAMpe.

The normalized surface concentration ($c_{i,\text{surf}}/c_{i,\text{max}}$) can be interpreted as the stoichiometry of the particle surface that determines the electrode potential.
The variation of the two limit values of $c_{i,\text{surf}}/c_{i,\text{max}}$ reflects the degradation effect on the stoichiometry ranges of both electrodes and the match between them.
We thus pick the first and last cycles and show in \cref{fig:sce1_sto}a the terminal voltage, PE potential, and NE potential versus the cell SoC of the discharge section (4th step in a cycle).
Corresponding to the first observation from \cref{fig:sce1}f, all three voltage curves of the last cycle start almost at the same points (fully-charged state) as their first-cycle counterparts, showing negligible differences between the first and last cycles.
The same starting points, in spite of the continuing PE degradation and the \LAMpe, can be explained as follows.
The cyclable lithium $M_\text{cyc,p}$ in the PE at the starting point (lower limit in \cref{fig:sce1}e) is zero for both the first and last cycles, and the concentration in the PE core takes the bottom value $c_\text{p,b}$ that equals the concentration $c_\text{s}$ of trapped lithium in the shell.
The progressive phase transition thus has no impact when the PE particle has zero cyclable lithium.

As discussed above, the \LAMpe results in earlier termination of the discharge.
This is further reflected in \cref{fig:sce1_sto}a by the increase of the SoC lower limit from 0.04 in the first cycle to 0.36 in the last cycle.
Since the NE has no degradation and its active material volume remains unchanged, the NE concentration and potential vary in the same pace with the cell SoC.
Therefore, the NE potential curve (versus the cell SoC) of the last cycle simply overlays the first-cycle curve at the high-SoC side (\cref{fig:sce1_sto}a), showing a decrease of the NE potential at the end of discharge from the first to the last cycle.
Accordingly, the NE particle stoichiometry at the end of discharge increases from 0.01 to 0.29 (\cref{fig:sce1_sto}b), corresponding to the second observation from \cref{fig:sce1}f.
The same lower cut-off value (\SI{2.8}{\volt}) is imposed to the terminal voltage---compare the solid and dashed purple lines (\cref{fig:sce1_sto}a); it follows that the PE potential is driven to be lower accordingly.
A lower PE potential indicates higher surface concentration and stoichiometry, from 0.94 to 1.0 as shown in (\cref{fig:sce1_sto}b).
In summary, \cref{fig:sce1_sto}b shows that the stoichiometry range of the NE particle during the discharge shrinks from $[0.81, 0.01]$ (the first cycle) to $[0.81, 0.29]$ (the last cycle), while the PE stoichiometry range expands from $[0.23, 0.94]$ to $[0.23, 1.0]$, leading to a re-match between the stoichiometry ranges.

We can conclude that the capacity fade in this scenario is exclusively caused by the \LAMpe by lifting the SoC lower limit, and that the stoichiometry range of the PE particle is expanded in response to the shrinkage of the NE stoichiometry range, caused by the \LAMpe.
We remark that, in spite of the null \LLIcyc, the \LLItot is always increasing, in pace with the phase transition, but it is not a key factor.
Rather, it is the \LLIcyc that matters, which is further discussed in scenario II.

\begin{figure}
	\centering
	\includegraphics{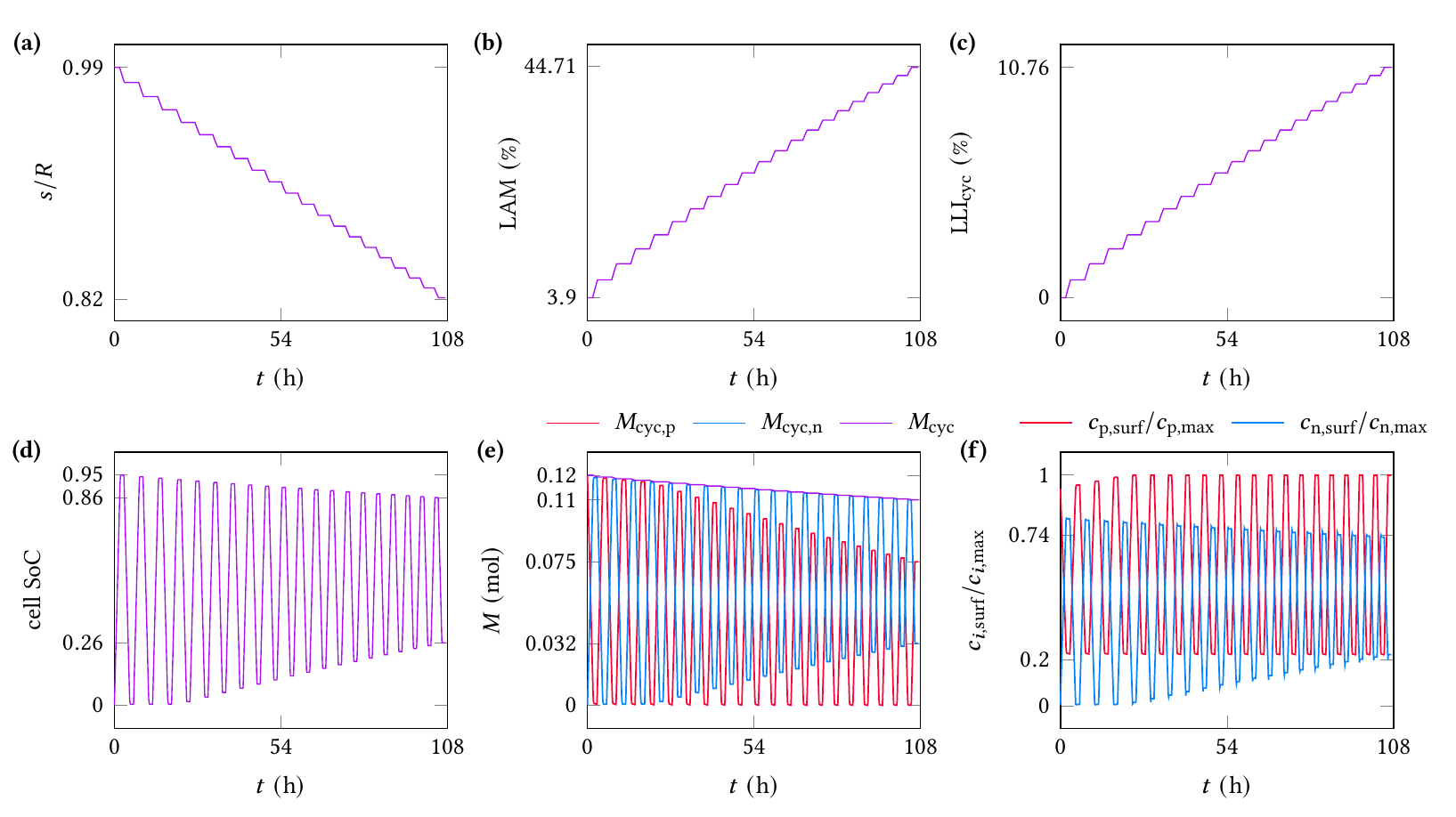}
	\caption{
	Scenario II: collective effects of loss of PE active material (\LAMpe) and loss of cyclable lithium (\LLIcyc) on cell performance in cyclic ageing tests.
	(a) core-shell phase boundary ($s/R$), (e) total cyclable lithium ($M_\text{cyc}$) in both electrodes, cyclable lithium in the PE ($M_\text{cyc,p}$) and NE ($M_\text{cyc,n}$), and (f) normalized particle surface concentration/stoichiometry ($c_{i,\text{surf}}/c_{i,\text{max}}$).
	}
	\label{fig:sce2}
\end{figure}

Compared to scenario I, scenario II has additional \LLIcyc that is achieved by setting $c_{\text{s}} > c_{\text{p,b}}$.
In \cref{fig:sce2}c, the \LLIcyc increases with time (and degradation), which is cross-checked by the decreasing total cyclable lithium $M_\text{cyc}$ in \cref{fig:sce2}e.
Note that we still keep $\rho = 0$ to have zero shell resistance and overpotential in this scenario.

Results of scenario II, as shown in \cref{fig:sce2}, are similar to those of scenario I in \cref{fig:sce1}, except for some differences caused by the introduced nonzero \LLIcyc.
First, the direct effect of \LLIcyc is the decrease of the SoC upper limit with cycle number from 0.95 to 0.86 as shown in \cref{fig:sce2}d, which can be explained by the same reasoning behind the constant upper limit at 0.95 in the case of zero \LLIcyc.
The interpretation is also supported by \cref{fig:sce1}e, in which the upper limit of cyclable lithium $M_\text{cyc,n}$ in the NE particle decreases at the same rate as the total cyclable lithium $M_\text{cyc}$.
Second, although the SoC lower limit also increases with cycle number due to the \LAMpe, the increase amplitude (0.26) in \cref{fig:sce2}d is smaller than that (0.36) in \cref{fig:sce1}d.
The smaller increase is exclusively due to the \LLIcyc, explained as follows.
The \LAMpe in scenario II is the same as in scenario I (44.71\% in \cref{fig:sce1}b and \cref{fig:sce2}b), and hence the decrease of the upper limit of the cyclable lithium $M_\text{cyc,p}$ in the PE, due to the \LAMpe, does not change from scenario I to scenario II (0.12 to 0.075 in both \cref{fig:sce1}e and \cref{fig:sce2}e).
Thus, the lifting effect of the \LAMpe on the lower limit of $M_\text{cyc,n}$ should be the same in both scenarios; the difference is that in scenario II the decrease in total cyclable lithium $M_\text{cyc}$ (i.e., the \LLIcyc) further drops the lower limit of $M_\text{cyc,n}$, thus counteracting the lifting effect of the \LAMpe.

The \LLIcyc shifts the SoC curve downwards---compare \cref{fig:sce1}d with \cref{fig:sce2}d.
However, the difference between the upper and lower SoC limits almost remains unchanged from scenario I to II, suggesting that no further capacity loss is caused by the \LLIcyc.
The capacity loss is dominated by the \LAMpe, but this conclusion is exclusive to the setting in our simulation.
Consider a scenario where the drop of $M_\text{cyc}$ in \cref{fig:sce2}e (the \LLIcyc) is larger than the decrease of the upper limit of $M_\text{cyc,p}$ due to the \LAMpe.
In such a case, the \LLIcyc effect may dominate, leading to the capacity loss.

\begin{figure}
	\centering
	\includegraphics{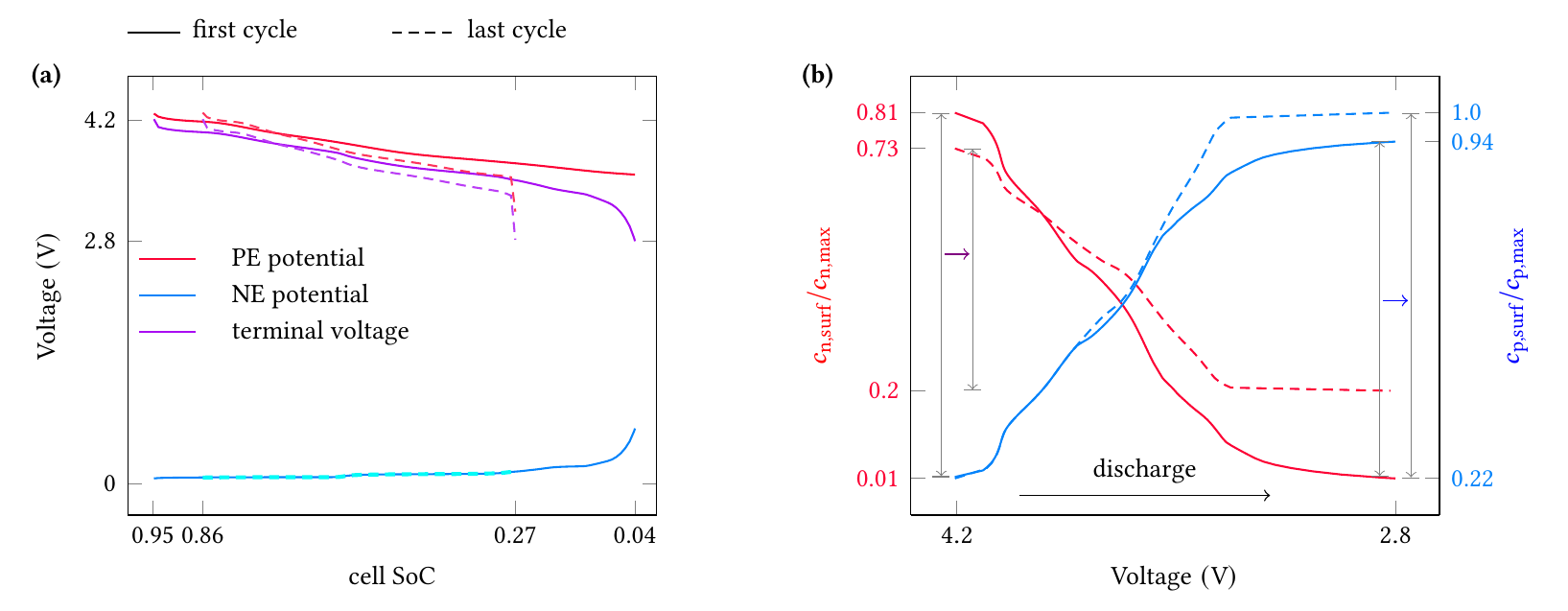}
	\caption{
		Scenario II: (a) terminal voltage, PE potential, and NE potential versus the cell SoC in the discharge step of the first and last cycles; (b) shrinkage of the NE stochiometry range and expansion of the PE stochiometry range during discharge.
	}
	\label{fig:sce2_sto}
\end{figure}

The downwards shifting of the SoC range in \cref{fig:sce2}d is also reflected in \cref{fig:sce2_sto}a.
Compared to \cref{fig:sce1_sto}a, the NE potential-SoC curve of the first cycle still overlap with part of the last-cycle curve, but slightly shifted to the low-SoC side.
The PE potential and terminal voltage of the last cycle are accordingly shifted towards lower SoCs.
The slight cell SoC decrease (0.95 to 0.86) from the first to the last cycle indicates that the NE is progressively less lithiated at the end of charge due to the \LLIcyc---see the change from 0.81 of the first cycle to 0.73 of the last cycle in \cref{fig:sce2_sto}b.
The NE potential thus becomes higher and consequently leads to a higher PE potential under the constraint of fixed upper cut-off voltage (\SI{4.2}{\volt}).
A higher PE potential at the end of charge further accelerates the PE degradation, acting as a positive feedback.
In \cref{fig:sce2_sto}a, the increases of the NE and PE potentials are relatively slight because the OCP used for the SiC NE in our simulation is basically flat at high lithiation levels.
As shown in \cref{fig:sce2_sto}b, the NE stoichiometry range is shrunk at both ends in this scenario: one by the \LAMpe and the other by the \LLIcyc.
Accordingly, the PE stoichiometry range is expanded at both ends, and a re-match between their stoichiometry ranges occurs.

We remark that the above-discussed model prediction of higher PE potential and accelerated PE degradation is confirmed by experimental ageing studies~\cite{Dose2020} of graphite/NMC811 full cells.
In Dose et al.~\cite{Dose2020}, the progressively less lithiated graphite is attributed to electrode slippage, which is also named stoichiometry drift and related to the stoichiometry range re-match in our simulations.
Our simulation results offer a possible explanation for the underlying cause of electrode slippage---the loss of cyclable lithium \LLIcyc.

\begin{figure}
	\centering
	\includegraphics{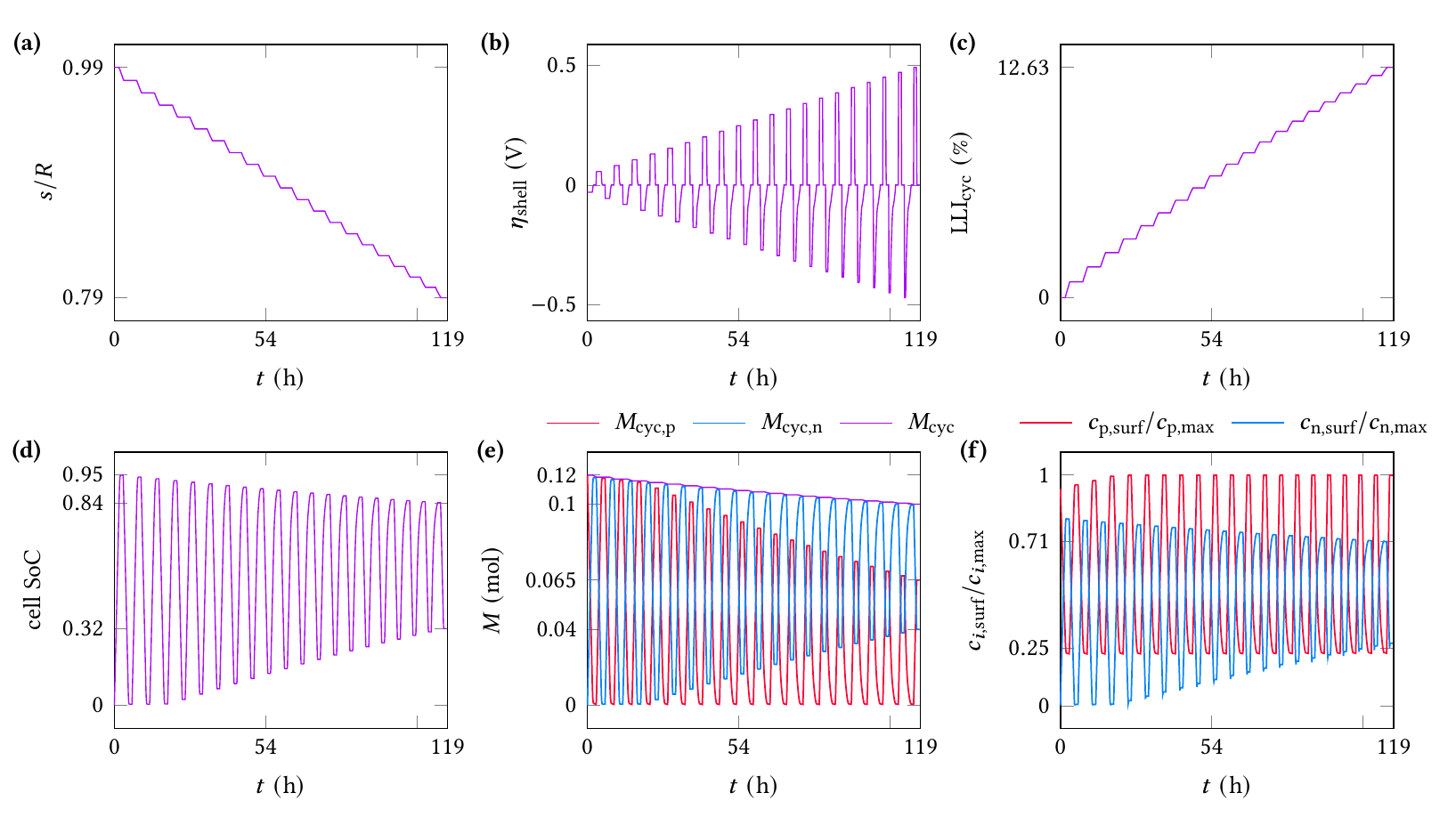}
	\caption{
	Scenario III: collective effects of loss of PE active material (\LAMpe), loss of cyclable lithium (\LLIcyc), and shell layer resistance on cell performance in cyclic ageing tests.
	(a) core-shell phase boundary ($s/R$), (b) overpotential ($\eta_\text{shell}$) across the shell layer, (e) total cyclable lithium ($M_\text{cyc}$) in both electrodes, cyclable lithium in the PE ($M_\text{cyc,p}$) and NE ($M_\text{cyc,n}$), and (f) normalized particle surface concentration/stoichiometry ($c_{i,\text{surf}}/c_{i,\text{max}}$).
	}
	\label{fig:sce3}
\end{figure}

Besides the \LAMpe and \LLIcyc, we further include the resistance of the shell layer and the resultant overpotential in scenario III.
As indicated by \cref{eq:shelloverpoten}, the shell overpotential depends on the current and the shell-layer thickness.
The variation of the shell overpotential $\eta_\text{shell}$ is shown in \cref{fig:sce3}b: its amplitude increases with cycle number because of the growing shell-layer thickness (or decreasing phase boundary location $s/R$ in \cref{fig:sce3}a); its frequency and variation pattern follows the dynamics of the current.

\begin{figure}
	\centering
	\includegraphics{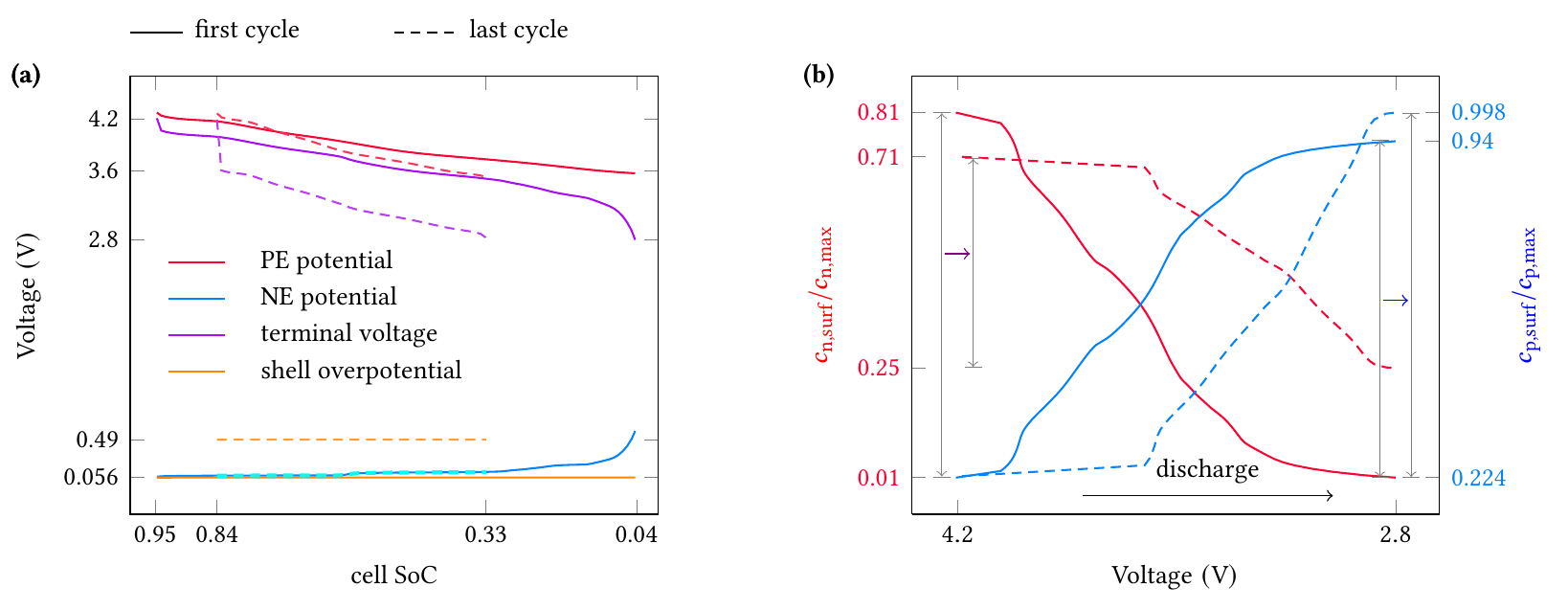}
	\caption{
		Scenario III: (a) terminal voltage, PE potential, and NE potential versus the cell SoC in the discharge step of the first and last cycles; (b) shrinkage of the NE stochiometry range and expansion of the PE stochiometry range during discharge.
	}
	\label{fig:sce3_sto}
\end{figure}

The shell-layer overpotential impacts the terminal voltage, as can be seen in \cref{eq:terminalvolt}.
During discharge, lithium enters the PE particle and the shell-layer overpotential $\eta_\text{shell}$ is negative according to \cref{eq:shelloverpoten}.
The difference incurred by the negative $\eta_\text{shell}$ is as follows.
During discharge, the terminal voltage drops from the upper cut-off voltage of \SI{4.2}{\volt}.
In \cref{fig:sce1_sto}a and \cref{fig:sce2_sto}a, the drop of the terminal voltage in the beginning of discharge is the same for the first and last cycles, as caused by reaction overpotential.
However, the terminal-voltage drop in \cref{fig:sce3_sto}a for the last cycle is up to \SI{0.6}{\volt}, mostly contributed by the shell overpotential (\SI{0.49}{\volt}).
The terminal voltage continues to stay at the low level during the whole discharge, diminishing the available power of the cell.

Furthermore, since the lower cut-off voltage is fixed, the extra shell overpotential leads to earlier termination of the discharge, i.e., the discharge ends at a higher SoC value (0.27 in \cref{fig:sce2_sto}a versus 0.33 in \cref{fig:sce3_sto}a), exacerbating the lifting effect by the \LAMpe.
This means the shell overpotential further narrows down the SoC range from the bottom side (end of discharge): the lower limit increases from 0.26 (\cref{fig:sce2}d) without the shell overpotential to 0.32 (\cref{fig:sce3}d) with overpotential.
Accordingly, the surface stoichiometry of the NE particle at the end of discharge increases from 0.2 (\cref{fig:sce2_sto}b) to 0.25 (\cref{fig:sce3_sto}b), suggesting a further shrinkage of the NE stoichiometry range from scenario II to III caused by the shell overpotential.
Note that the minor difference in the SoC lower limits between \cref{fig:sce3_sto}a (0.33) and \cref{fig:sce3}d (0.32) is because the voltage control following the discharge slightly pulls down the SoC.
The same phenomena and reasoning apply to scenario II in \cref{fig:sce2_sto}a and \cref{fig:sce2}d.

The SoC upper limit (0.84) in \cref{fig:sce3}d is also lower than that (0.86) in \cref{fig:sce2}d.
This small drop is mainly caused by a greater \LLIcyc---12.63\% in \cref{fig:sce3}c vs 10.76\% in \cref{fig:sce2}c.
During charge, the shell overpotential makes the cell further away from equilibrium when the upper cut-off voltage is reached, leading to a longer voltage control afterwards with degradation occurring.
The longer the degradation lasts, the higher the \LLIcyc value and the lower the $s/R$ value (0.82 $\to$ 0.79 from scenario II to III).
The longer voltage control agrees with the fact that a longer time (119 hours in \cref{fig:sce3}) to complete 20 cycles in scenario III than the 108 hours (\cref{fig:sce2}) used in scenario II.
In summary, the extra shell overpotential leads to further shrinkage of the cell SoC range (\cref{fig:sce2_sto}a vs \cref{fig:sce3_sto}a) and shrinkage of the NE stoichiometry range (\cref{fig:sce2_sto}b vs \cref{fig:sce3_sto}b) from both ends.
The shell overpotential also depresses the cell power by diminishing the terminal voltage.

\begin{figure}
	\centering
	\includegraphics{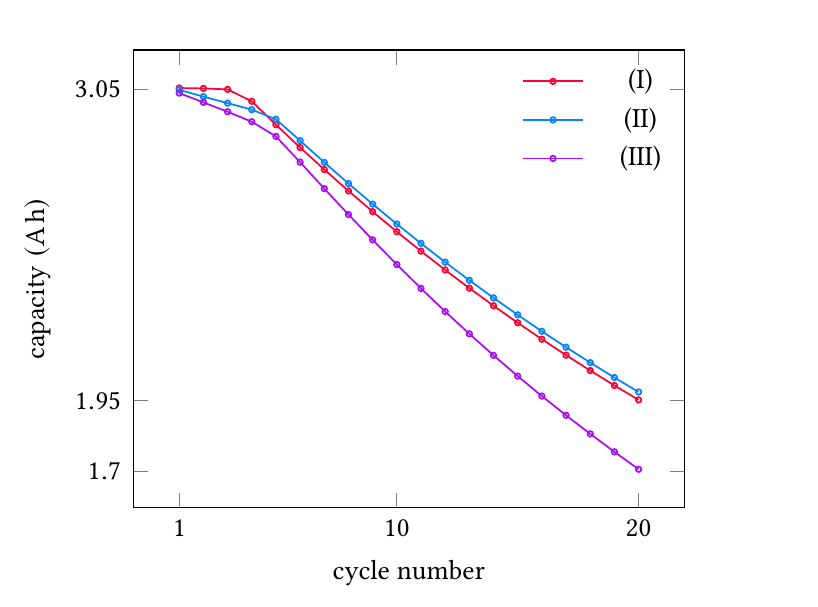}
	\caption{
		Discharge-capacity fade for scenario I considering the \LAMpe, scenario II considering the \LAMpe and the \LLIcyc, and scenario III considering the \LAMpe, the \LLIcyc, and shell-layer resistance.
	}
	\label{fig:cap_fade}
\end{figure}

Finally, we plot the discharge-capacity fade for all three scenarios in \cref{fig:cap_fade}.
The difference between scenario I and II is negligible because, as discussed above, the \LAMpe effect is dominant over the \LLIcyc effect such that the additional \LLIcyc does not result in extra capacity loss.
The capacity in scenario III drops faster because of the extra shell layer overpotential causing further shrinkage of the cell SoC range.


\subsection{Degradation inhomogeneity in through-cell direction}
\label{sec:inhomogeneity}

In the previous section, we present the degradation in the average sense and disregard its variation in the direction of electrode thickness.
Here we present the degradation inhomogeneity across the electrode thickness caused by a constant-current charge at a 1\,C rate followed by a voltage control and a half-hour rest.
In this case, the particle degradation model is implemented within the DFN model in PyBaMM.

\begin{figure}
	\centering
	\includegraphics{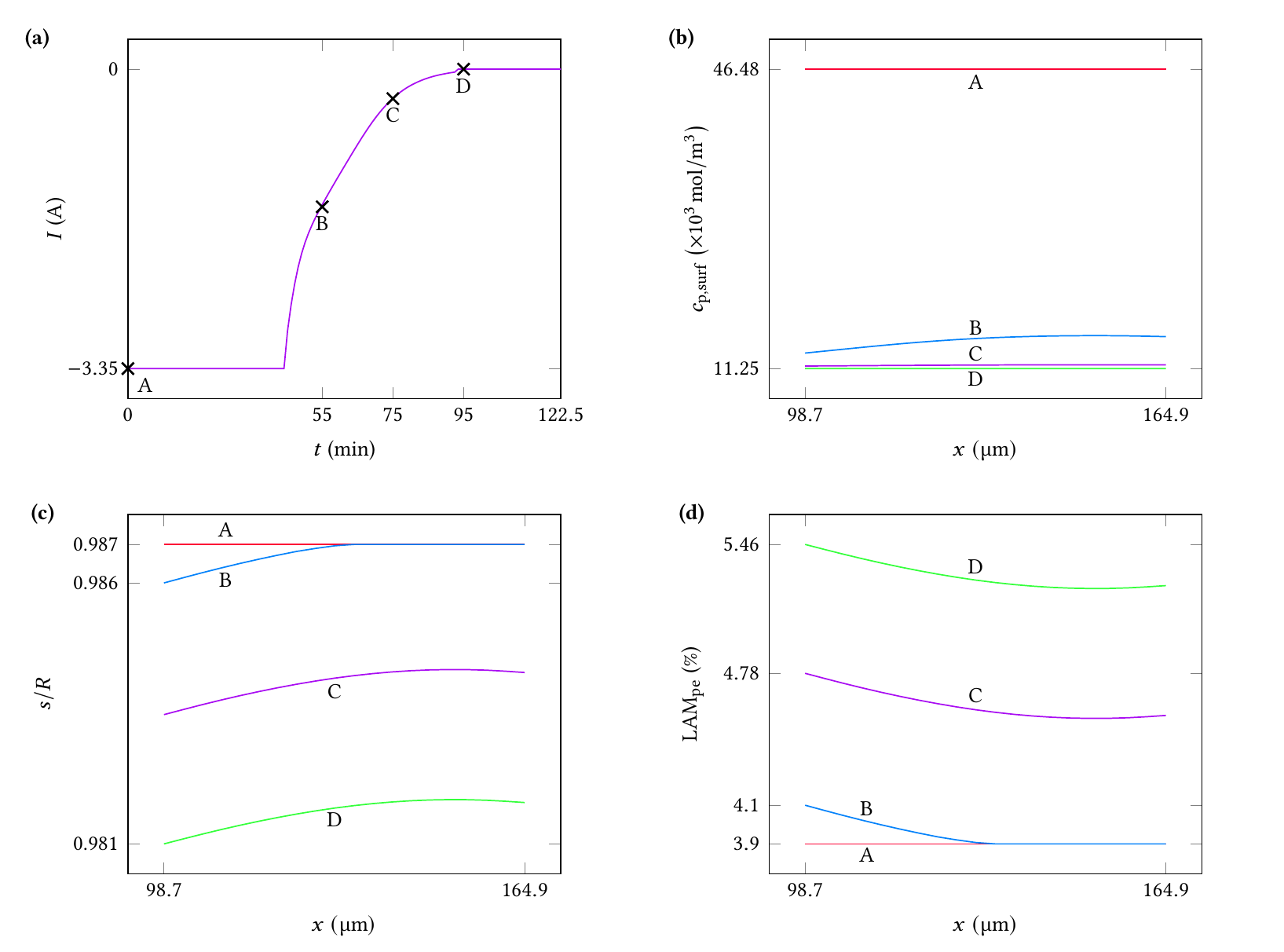}
	\caption{
		Degradation inhomogeneity along the electrode thickness direction in a numerical test of cell charge followed by voltage control and rest.
		The current $I$ for the test protocol is shown in subplot (a) as a function of time in minutes.
		(b) Positive core surface concentration $c_\text{p,surf}$, (c) phase boundary location $s/R$, and (d) loss of PE active material \LAMpe at four time instants corresponding to four points A--D in (a).
		The PE is represented by $98.7 \leq x \leq \SI{164.9}{\micro\meter}$ in the thickness direction.
	}
	\label{fig:dfn}
\end{figure}

The current variation, in accordance with the specified protocol, is shown in \cref{fig:dfn}a.
In particular, the current decreasing in magnitude results from the voltage control.
During the procedure, we pick four time instants, corresponding to points A--D in \cref{fig:dfn}a, and show the positive core surface concentration $c_\text{p,surf}$, phase boundary location $s/R$, and the loss of PE active material \LAMpe in subplots b--d, respectively.
The three quantities in b--d not only vary with the time but also depend on the depth $x$ through the PE relative to the current collector at the NE side ($\SI{98.7}{\micro\meter}$ is the thickness of the NE and separator).
At the initial time (point A), the lithium concentration is set to be uniform across the electrode and inside the core of each PE particle, and thus the core surface concentration is constant in the $x$ direction (\cref{fig:dfn}b).
The phase boundary location (\cref{fig:dfn}c) is also set to be constant initially in the whole electrode.
During the constant-current charge and voltage control, lithium are removed from the PE and inserted into the NE, leading to a continuous decrease of the PE core (surface) concentration.
In particular, the surface concentration is lower at the PE-separator boundary, and this is because the interfacial current density is larger for PE particles closer to the separator where the resistance for current flow is smaller \cite{Geng2021}.
By the time at point B, the PE core surface concentration has dropped dramatically to the extent that the surface concentration at the separator side ($x = \SI{98.7}{\micro\meter}$ in \cref{fig:dfn}b) is lower than the phase-transition threshold value.
Hence, the phase boundary moves inwards for particles close to the separator; the lower the surface concentration, the larger extent to which the phase transition occurs and the lower the $s/R$ value.
The surface concentrations of particles close to the current-collector side are still above the phase-transition threshold value, and thus there is yet no phase transition at higher $x$ values in \cref{fig:dfn}c.
After point B, the cell charge gradually fades until the current vanishes.
Accordingly, we just observe a slight decrease of the surface concentration from B to C and D in \cref{fig:dfn}b.
However, the phase boundary location $s/R$ continues to decrease (B$\to$C$\to$D) as the surface concentration remains low.
The low surface concentration enables the phase transition to proceed across the whole electrode, and the speed of phase change (time rate of $s/R$ in \cref{eq:pbmoving}) is basically uniform in the $x$ direction.
Therefore, $s/R$ as a function of $x$ shifts downwards as a whole in \cref{fig:dfn}c.
The \LAMpe at the four time instants in \cref{fig:dfn}d shows similar profiles and behavior to the phase boundary $s/R$ because of the definition (\ref{eq:lam}).

\begin{figure}
	\centering
	\includegraphics{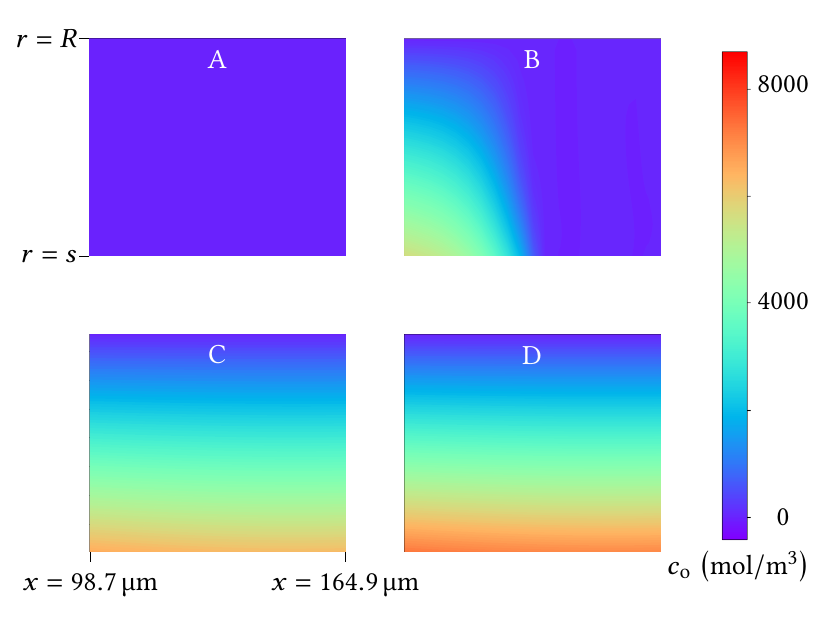}
	\caption{
		Oxygen concentration $c_\text{o}$ contour at four time instants ($t = 0$, $55$, $75$, \SI{95}{\minute}) corresponding to the four points A--D marked in \cref{fig:dfn}a, respectively.
	}
	\label{fig:oxygenconc}
\end{figure}

\cref{fig:oxygenconc} shows the oxygen concentration contours at the four chosen time instants.
The oxygen concentration is solved in the shell phase of each PE particle across the entire electrode thickness.
The horizontal axis ($x$) denotes the electrode thickness direction, and the vertical axis ($r$) represents the radial direction of a particle.
Note that the shell thickness ($R-s$) varies with time and the coordinate $x$, and that the heights of the four subplots do not represent the actual shell thickness but are scaled accordingly.
Initially at point A, the oxygen concentration is uniform in both directions and stays null.
At point B, as the phase transition occurs mostly at the separator side, the oxygen stored in the cores is released into the shells at lower $x$ values, forming the concentration hot point in the lower-left corner.
Once generated at the core-shell boundary, the oxygen diffuses through the shell, as indicated by the concentration decay in the $r$ direction.
The concentration decay in the $x$ direction is due to the lower levels of phase transition at locations away from the separator (\cref{fig:dfn}c).
With phase transition spreading from the left-hand side (B) to the whole electrode (C and D), the generated oxygen is present throughout the whole electrode, and the longer the phase transition undergoes, the higher value of the oxygen concentration (C vs D).
Meanwhile, the oxygen concentration gradient remains in the $r$ direction due to the boundary condition of zero oxygen at the shell surface.

\section{Conclusion}

We developed a shrinking-core particle model to describe the degradation mechanism of phase transition for high-nickel PEs and presented a two-step strategy to understand the degradation effects.
We have shown how the progress of phase transition directly results in the primary degradation modes of loss of PE active material of (\LAMpe), loss of lithium inventory (\LLItot and \LLIcyc), and additional shell-layer resistance (\cref{sec:storage}); the primary degradation modes then deteriorate the cell performance in terms of capacity and power (\cref{sec:cyclicaging}).

It is found that the \LAMpe lifts the lower limit of the SoC range by terminating the discharge earlier, while the \LLIcyc suppresses the upper limit of the SoC range; in our simulated cases, the \LAMpe is the dominant factor contributing to the capacity loss.
The shell-layer overpotential shrinks the SoC range from both ends, leading to a power reduction and further capacity fade under a fixed-voltage-window operation.
The \LAMpe and \LLIcyc narrow down the stoichiometry range of the NE and consequently expand the PE stoichiometry range due to the fixed-voltage window.
Due to the flat NE open circuit potential at high lithiation, the increase of PE potential at the end of charge is slight; for a NE with steeper potential curve, we expect a noticeable PE potential increase caused by the \LLIcyc, leading to a positive feedback and accelerating the PE degradation.

We find that it is the \LLIcyc, not \LLItot, that impacts the cell performance, and thus we suggest to explicitly differentiate the cyclable lithium from the total lithium in the calculation of LLI.
\LLItot always occurs during the phase transition, but this is not the case for \LLIcyc; the \LLIcyc depends on the concentration of lithium ($c_\text{s}$) trapped in the degraded materials, which is assumed as a constant in our model.
Thus, the calibration of parameter $c_\text{s}$ against experiments is key to identifying the LLI effects.

The study of LAM and LLI effects on capacity fade has offered insights into experimental diagnosis of battery degradation modes.
Specifically, we can differentiate the contribution to degradation by LAM from that by LLI according to the cell SoC variation curve and especially its change pattern.
With the primary degradation modes captured, the model can qualitatively reproduce experimentally-observed phenomena (e.g., capacity fade and stoichiometry drift).
However, challenges still remain in model calibration and validation against specifically-designed experiments.

\section*{Acknowledgements}
The research leading to these results has received funding from the Innovate UK through the WIZer Batteries project (grant number 104427) and EPSRC Faraday Institution Multi-Scale Modelling project (EP/S003053/1, grant number FIRG003).
The support from PyBaMM development team is appreciated in our model implementation within PyBaMM.

\appendix

\section{Numerical implementation in PyBaMM}

The particle degradation model is implemented as a submodel in PyBaMM~\cite{Sulzer2021}, which is an open-source python programming package aimed at solving physics-based electrochemical models (differential algebraic equations).
The submodel can then be called by the Single Particle Model and Doyle-Fuller-Newman model available in PyBaMM, in place of the original particle model that only considers fickian diffusion.
The nondimensionalization of equations and boundary conditions and implementation (codes) can be found in the public PyBaMM repository.

Regarding the degradation particle model, there are two numerical challenges to address.
The first is that the computational domains of the core and shell are changing with time due to the moving phase boundary.
To fix this problem, we follow the same numerical trick as in Refs.~\cite{Zhang2007,Ghosh2021} and define two new spatial variables $\eta$ and $\chi$ for the core and shell, respectively:
\begin{align} \label{eq:varichange}
	\eta = \frac{r}{s}, \quad
	\chi = \frac{r-s}{R-s},
\end{align}
where $r$ is the spatial coordinate in the radial direction, $s$ denotes the phase boundary location, and $R$ is the particle radius (see \cref{fig:core_shell_sche}).
Now the computational domains of the core and shell both reduce to $\eta, \chi \in [0, 1]$.

\begin{figure}
	\centering
	\includegraphics{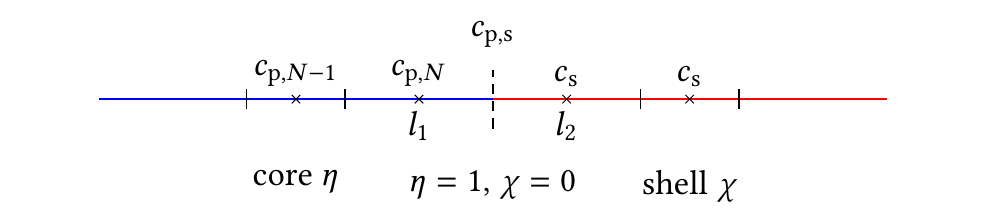}
	\caption{
	Schematic of discretization of the moving phase boundary condition~\eqref{eq:interface_masscons_4}.
	Parameter $c_\text{p,s}$ denotes the value of concentration $c_\text{p}$ at the core-shell phase boundary, and $N$ is the number of cells after discretization using the finite volume method. 
	}
	\label{fig:discretize}
\end{figure}

The second is the discretization of the Robin-type boundary condition~\eqref{eq:interface_masscons_4}.
Referring to \cref{fig:discretize}, the discretized version can be expressed as
\begin{align} \label{eq:interface_masscons_5}
	\dot{s} \qty(c_{\text{s}} - c_\text{p,s}) -
	D_\text{p} \frac{2}{s} \frac{ c_\text{p,s} - c_{\text{p,}N} }{l_1} -
	\qty(R/s)^2 \frac{j_\text{p}}{F} = 0.
\end{align}
The boundary value $c_\text{p,s}$ is solved from \cref{eq:interface_masscons_5} and used to express the boundary flux term (the middle term), and then the flux term is applied in the same way as the normal Neumann boundary condition.
The same procedure applies to the boundary condition \eqref{eq:interface_oxygen_2} for oxygen diffusion.

\addcontentsline{toc}{section}{References}
\bibliography{./myrefs.bib}

\end{document}